\documentclass[preprint,12pt]{elsarticle}

\usepackage{amssymb}
\usepackage{amsmath}
\usepackage{graphicx}% Include figure files
\usepackage{dcolumn}% Align table columns on decimal point
\usepackage{bm}% bold math
\usepackage{orcidlink}
\usepackage{physics}
\usepackage[utf8]{inputenc}
\usepackage{hyperref}
\begin{document}

\begin{frontmatter}

%% Title, authors and addresses

%% use the tnoteref command within \title for footnotes;
%% use the tnotetext command for theassociated footnote;
%% use the fnref command within \author or \affiliation for footnotes;
%% use the fntext command for theassociated footnote;
%% use the corref command within \author for corresponding author footnotes;
%% use the cortext command for theassociated footnote;
%% use the ead command for the email address,
%% and the form \ead[url] for the home page:
%% \title{Title\tnoteref{label1}}
%% \tnotetext[label1]{}
%% \author{Name\corref{cor1}\fnref{label2}}
%% \ead{email address}
%% \ead[url]{home page}
%% \fntext[label2]{}
%% \cortext[cor1]{}
%% \affiliation{organization={},
%%             addressline={},
%%             city={},
%%             postcode={},
%%             state={},
%%             country={}}
%% \fntext[label3]{}

\title{Dynamics of localised states in the stochastic discrete nonlinear Schr\"odinger equation} %% Article title

%% use optional labels to link authors explicitly to addresses:
%% \author[label1,label2]{}
%% \affiliation[label1]{organization={},
%%             addressline={},
%%             city={},
%%             postcode={},
%%             state={},
%%             country={}}
%%
%% \affiliation[label2]{organization={},
%%             addressline={},
%%             city={},
%%             postcode={},
%%             state={},
%%             country={}}

\date{June 30, 2025}

\author[first]{Mahdieh Ebrahimi\corref{c1}}
\cortext[c1]{Corresponding author}
\ead{mahdieh.ebrahimi@pkm.tu-darmstadt.de}
\affiliation[first]{organization={Institute for Condensed Matter Physics, TU Darmstadt},%Department and Organization
            addressline={Hochschulstrasse 6}, 
            city={Darmstadt},
            postcode={D-64289}, 
            %state={Hessen},
            country={Germany}}

\author[first]{Barbara Drossel}
\ead{barbara.drossel@pkm.tu-darmstadt.de}
%\affiliation[second]{organization={Institute for Condensed Matter Physics, TU Darmstadt},%Department and Organization
%            addressline={Hochschulstrasse 6}, 
%            city={Darmstadt},
%            postcode={D-64289}, 
%            %state={Hessen},
%            country={Germany}}

\author[third]{Wolfram Just}
\ead{wolfram.just@uni-rostock.de}
\affiliation[third]{organization={Institute of Mathematics, University of Rostock},
            addressline={Ulmenstra\ss e 69}, 
            city={Rostock},
            postcode={D-18057}, 
            %state={},
            country={Germany}}

%% Abstract
\begin{abstract}
We revisit aspects of dynamics and stability of localised states in the deterministic and stochastic discrete nonlinear Schr\"odinger equation. By a combination of analytic and numerical techniques, we
show that for deterministic motion localised initial conditions disperse
if the strength of the nonlinear part drops below a threshold and that localised states are unstable in a noisy environment. As expected, the constants of motion in the nonlinear Schr\"odinger equation play a crucial role. An infinite temperature state emerges when multiplicative noise is applied, while additive noise yields unbounded dynamics since conservation of normalisation is violated.
\end{abstract}

%%Graphical abstract
%\begin{graphicalabstract}
%\includegraphics{grabs}
%\end{graphicalabstract}

%%Research highlights
%\begin{highlights}
%\item Research highlight 1
%\item Research highlight 2
%\end{highlights}

%% Keywords

\end{frontmatter}

%% Add \usepackage{lineno} before \begin{document} and uncomment 
%% following line to enable line numbers
%% \linenumbers

%% main text
%%

%% Use \section commands to start a section
\section{\label{sec:1}Introduction}

Since Holstein introduced the Discrete Nonlinear Schr\"odinger Equation (DNSE) model in 1950 \cite{HOLSTEIN1959325} to describe polaron motion in molecular crystals, it has been widely applied to various phenomena, including wave propagation in nonlinear arrays \cite{HENNIG1999333}, Bose-Einstein condensation in optical lattices \cite{PhysRevLett.86.2353,Franzosi_2011}, and energy transport in biomolecules \cite{10.1093/oso/9780198528524.001.0001}. As a nonintegrable equation, the DNSE exhibits remarkable features such as chaotic trajectories, periodic orbits, and breather solutions, which are spatially localised, time-dependent excitations (see, for instance, the review \cite{FlWi_PR98} or the recent monograph \cite{kevrekidis_discrete_2009}). 

The DNSE is one of the few paradigmatic model systems where fundamental aspects of statistical mechanics and high-dimensional Hamiltonian dynamics have been studied in considerable detail, including non-standard features such as
negative-temperature states (see e.g. the review \cite{BaIuLiVu_PR21}).
In addition to the energy, the DNSE admits another, nontrivial, macroscopic constant of motion, which can be either identified with the normalisation condition of the state in phase space or the particle number. 
These two constants of motion are crucial
to understand the features of the model.
The statistical mechanics of such models have been coined in
\cite{rasmussen_statistical_2000, rasmussen_discrete_2000, RuNe_PRL01}. 
where a Gibbs measure based 
on the two conserved quantities, energy and particle number, has been analysed.
However, this measure breaks down at negative temperatures due to the ill-defined nature of the grand canonical ensemble, and this
breakdown is related to the occurrence of breather solutions
\cite{rumpf_simple_2004, Rumpf_2007, rumpf_transition_2008, RUMPF20092067}.
The transition to negative temperature states has much in common with first order phase transitions and the creation of thermodynamically metastable states. 
While the energy is predominantly contained in a few localised structures, the
entropy of the system is largely determined by low-amplitude parts of the solutions.
%based on entropic arguments, that starting from a negative temperature state an %increasing amount of energy will become concentrated in breathers while $\beta = %1/k_BT$ increases towards 0. 
%such that the system's final state should converge to a single high-amplitude %breather state at infinite temperature. 
Detailed numerical studies showed that the dynamics involves extremely slow relaxation processes \cite{PhysRevLett.122.084102,iubini_coarsening_2014}. These findings suggest that the system evolves towards a stationary state with a finite breather density and a microcanonical temperature that remains negative \cite{PhysRevE.70.066610, Iubini_2013, BaIuLiVu_PR21}.

%%While studies of stochastic versions of the DNSE in the physics context seem to be scarce,
The impact of noise on the DNSE and on the spatially continuous nonlinear Schr\"odinger equation has been extensively studied from a rigorous mathematical point of view in considerable detail \cite{de_bouard_stochastic_1999,Carlen_2019,carlen_exponential_2016,chatterjee_invariant_2014,hannani_stochastic_2023,chatterjee_probabilistic_2012}.  
While the energy, due to noise, is not any longer
a constant of motion, the stochastic DNSE still preserves the particle number.
Among others, the existence of a unique invariant Gibbs measure has been established for a multiplicative noise model with a phenomenological damping term
\cite{hannani_stochastic_2023}. 
Again, these studies emphasise that the conservation of normalisation is a crucial feature of the DNSE. 

Here, we want to focus on the behaviour of localised solutions in a noisy environment. Since we view the DNSE as an effective equation of motion, 
covering for instance the evolution of a many-particle system within a mean field description,
we use a formulation where we do not eliminate the parameter which measures the strength of the nonlinear part (see eq.(\ref{eqofmotion})), in contrast to the standard form normally used in studies on the statistical mechanics of the model
(see eq.(\ref{original})). While both descriptions are equivalent from a dynamical systems point of view, there are
differences when the thermodynamic limit is considered. Our approach
is more amenable to a system of finite size and may miss features which are relevant in the thermodynamic limit. Having said that, this approach enables us to map negative temperature states to positive temperature states and vice versa by changing the sign of the nonlinearity, normally called the
focusing or defocusing case of the nonlinear Schr\"odinger equation.
In section \ref{sec:2} we summarise some basic properties of the equations of motion and of the properties and stability of breather states, covering the focusing as well as the defocusing case. While these features are well established in the existing literature, we cover, in addition, the evolution of localised initial conditions, which tend to delocalise when the strength of the nonlinear part falls below a certain threshold value. Section \ref{sec:3} is devoted to the impact of noise on breather states. We will analyse two cases: multiplicative noise, which respects conservation of normalisation of the DNSE, and additive noise, which violates such a conservation law. As expected, localised states do not prevail in such noisy environments, but the former case still leads to a stationary state, emphasising the relevance of conserved quantities for the dynamics of the DNSE. 
Our present studies with noise correspond to an infinite temperature condition imposed on the dynamical system, so that suitable damping terms are needed to enter the
more interesting finite temperature domains. Some of the implications will be addressed in the conclusion, section \ref{sec:4}.
To keep our account self-contained we also include details about the numerical
integration schemes for Hamiltonian dynamics in the appendix.

\section{The Discrete Nonlinear Schr\"odinger Equation (DNSE)}\label{sec:2}
\subsection{Symmetries and Transformations} \label{sec:2.1}

We employ a version of the one-dimensional DNSE with a tunable coefficient $\alpha$ 
\begin{align}
    \label{eqofmotion}
    \dot{c}_n(t) &= i \Big( 2c_n(t) - c_{n-1}(t) - c_{n+1}(t) \nonumber\\
    &- \alpha |c_n(t)|^{2}c_n(t)\Big) \, .
\end{align}
Here, $c_n$ represents the complex phase or amplitude at each site $n$, 
where we assume a system of size $L$ imposing periodic boundary conditions
$c_n=c_{n+L}$. The parameter $\alpha$ controls the strength of the nonlinearity.
The choice $\alpha>0$ leads to a so-called focusing and the choice 
$\alpha<0$ to a so-called  defocusing nonlinear Schr\"odinger equation.
The equation of motion, eq.~\eqref{eqofmotion}, constitutes a
Hamiltonian system where $\{ c_n, \bar{c}_n \}$ are canonical conjugate variables,  and the Hamiltonian itself is given by
\begin{equation}
    \label{Hamiltonian}
    H = \sum_{n=0}^{L-1} \left(2|c_n|^2 - c_n \bar{c}_{n-1} - c_n \bar{c}_{n+1} -
    \frac{\alpha}{2}|c_n|^4 \right) \, . 
\end{equation}
Using the appropriate Poisson bracket
\begin{equation}\label{eq2.1}
\{F,G\}=i \sum_n \left(\frac{\partial F}{\partial \bar{c}_n}
\frac{\partial G}{\partial c_n} -
\frac{\partial F}{\partial c_n}
\frac{\partial G}{\partial \bar{c}_n}\right) \, ,
\end{equation}
the DNSE can be cast into the form of canonical equations of motion
\begin{equation}\label{eq2.2}
\dot{c}_n(t)=i\frac{\partial H}{\partial \bar{c}_n}=\{H,c_n\}=
i\mathbf{L} c_n 
\end{equation}
where $\mathbf{L}$ denotes the Liouville operator 
$i\mathbf{L}\cdot=\{H,\cdot\}$. Incidentally, eq.~\eqref{eq2.1} imposes a sign convention
which also determines the sign of the Hamiltonian (\ref{Hamiltonian}).

It is a key feature of the nonlinear Schr\"odinger equation that it admits
two conserved quantities, namely,
the system's total energy $E = H$ and the normalisation of the wave function 
or the particle number
\begin{equation}\label{eq2.3}
N = \sum_{n=0}^{L-1} |c_n|^2\, ,
\end{equation}
as can be easily derived from $\{H,N\}=0$. In our setup, we chose $N=1$ and let
the total energy $E$ and the on-site potential strength $\alpha$ be the two independent parameters of the dynamics.

The DNSE is often used in non-dimensional units 
(see e.g. \cite{rumpf_simple_2004,BaIuLiVu_PR21}) where it takes the form
\begin{equation}
    \label{original}
    i \dot{C}_n = C_{n+1} + C_{n-1} + |C_n|^{2}C_n \, .
\end{equation}
This form is obtained from eq.~\eqref{eqofmotion} by performing
the phase transformation $c_n \mapsto \exp(- i 2 t) c_n$ and applying the scaling
$c_n\mapsto \sqrt{|\alpha|} c_n$ in the focusing case $\alpha>0$. In the defocusing case, $\alpha<0$, this procedure results in
eq.~\eqref{original}, with a negative sign of the cubic term. For systems with
an even number of lattice sites such an equation with negative cubic term can be transformed into the normal form, eq.~\eqref{original}, by using $c_n \mapsto (-1)^n \bar{c}_n$. Thus, for the DNSE there is no fundamental difference between the focusing and the defocusing case from a dynamical systems point of view.
The two conserved quantities of eq.~\eqref{original} are the energy $\tilde{H}$ and the particle number $\tilde{N}$
\begin{align}\label{eq2.4}
\tilde{H}&=\sum_{n=0}^{L-1}\left( C_n \bar{C}_{n+1}+C_n \bar{C}_{n-1} 
+\frac{1}{2} |C_n|^4 \right), \nonumber\\
\tilde{N}&=\sum_{n=0}^{L-1} |C_n|^2
\end{align}
which are considered to be the two parameters of the system. The dynamical 
behaviour of the model can be captured by phase diagrams which are presented in
terms of these quantities \cite{rumpf_simple_2004,BaIuLiVu_PR21}.
Using the above transformations between the $C_n$ and $c_n$, the values of these conserved quantities can be
expressed in terms of eqs.~\eqref{Hamiltonian} and \eqref{eq2.3} as
\begin{equation}\label{eq2.5}
\tilde{H}=\alpha(2N-H)=\alpha(2-E), \qquad \tilde{N}=|\alpha| N= |\alpha | \, .
\end{equation}
In the following, we present our results in terms of eq.~\eqref{eqofmotion} with $N=1$ and the two parameters $E$ and $\alpha$, while eq.~\eqref{eq2.5} can be used to translate those results
to the non-dimensional version eq.~\eqref{original}. Note that due to the normalisation condition the energy $E$ is not extensive, so that $-\alpha/2\leq E \leq 4-\alpha/(2L)$ if $\alpha>0$, and $-\alpha/(2L)\leq E \leq 4-\alpha/2$ if $\alpha<0$. 

\subsection{Breathers and their Stability} \label{sec:2.2}

\begin{figure}[!ht]
    \centering
    \includegraphics[width=0.8\linewidth]{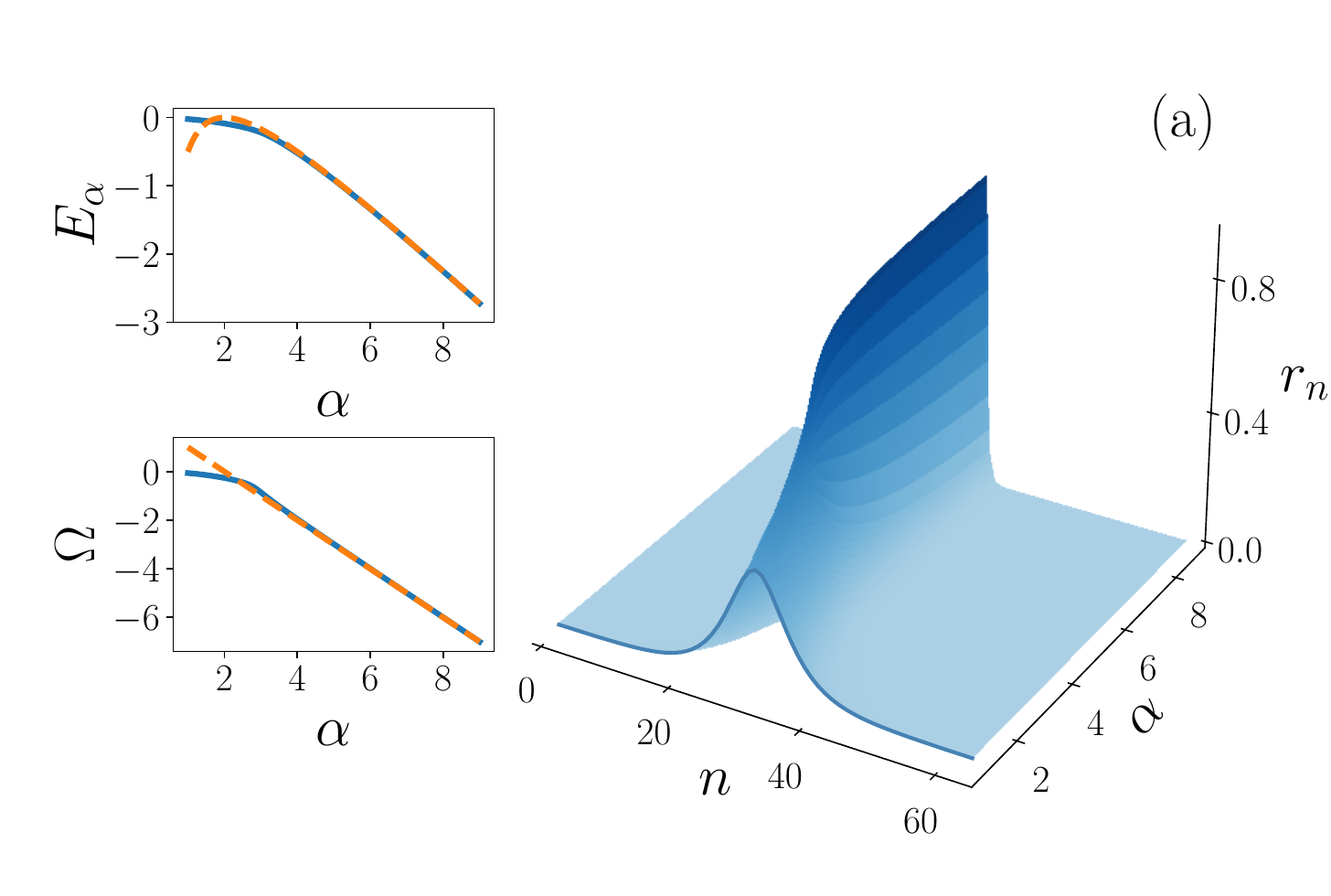}
    \includegraphics[width=0.8\linewidth]{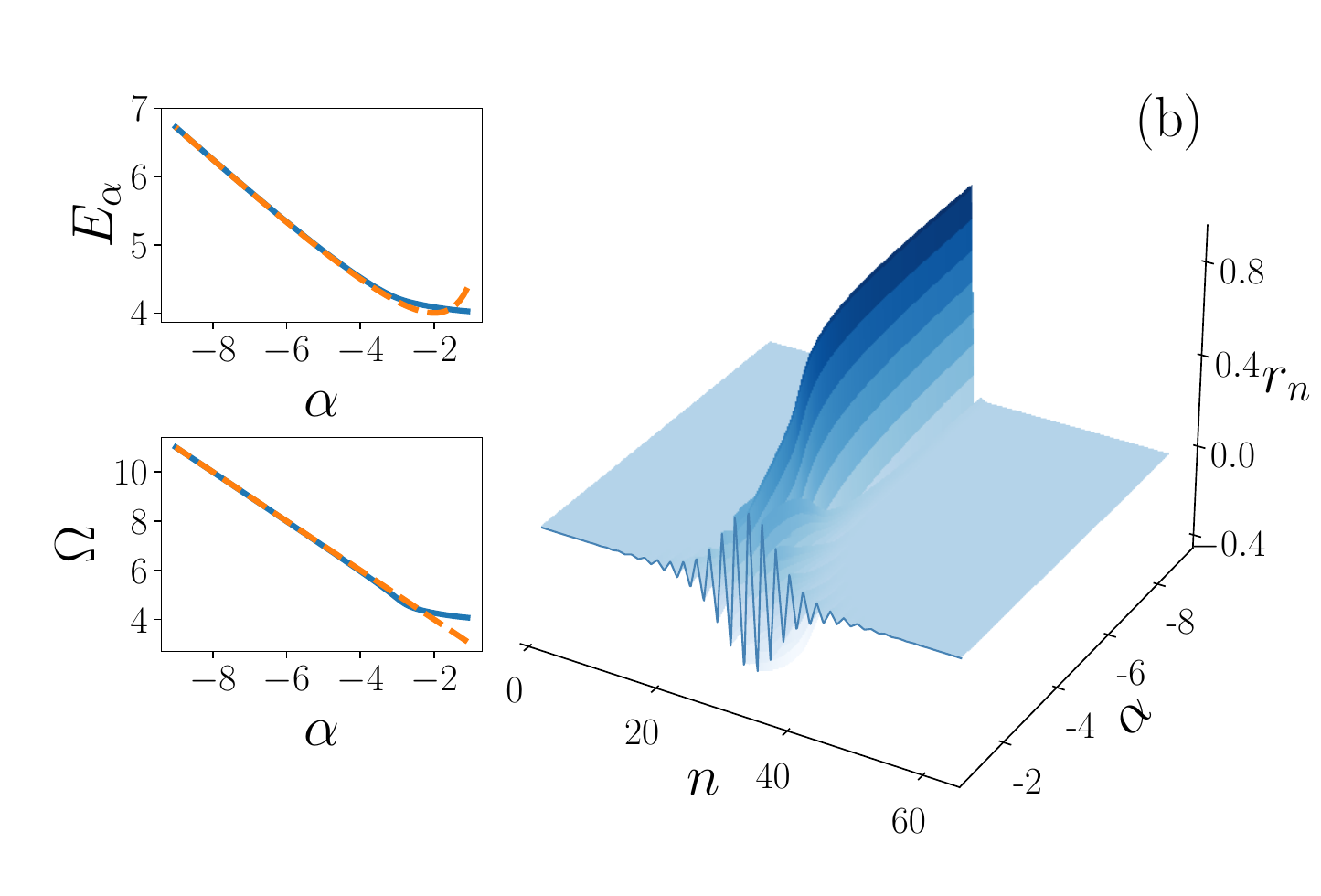}
\caption{
(a) Spatial profile $r_n$ of a breather in dependence on $\alpha$ for (a) the focusing  and (b) the defocusing DNSE, obtained from
eq.~\eqref{eq2.6} for a system of size $L=64$. 
The insets show the respective dependence of the
energy $E$ and of the breather frequency $\Omega$ on $\alpha$ (blue, solid lines), as well
as the leading order asymptotic expression in eq.~\eqref{eq2.8} (amber, broken line). 
\label{fig1}}
\end{figure}

The nonlinear Schr\"odinger equation is one of the paradigmatic
model systems where time-dependent spatially localised solutions, so-called breathers, have been studied in considerable detail \cite{johansson_existence_1997}. Here we summarise a few of the main well-known features. In the context of eq.~\eqref{eqofmotion} the simplest type of breather solution has the form $c_n(t)=\exp(i\Omega t) r_n$, where
the frequency $\Omega$ determines the period of the breather and $r_n$ the real-valued spatially localised shape. With eqs.~\eqref{eqofmotion} and \eqref{eq2.3} we obtain the nonlinear eigenvalue problem
\begin{equation}\label{eq2.6}
0=(2-\Omega-\alpha r_n^2)r_n-r_{n-1}-r_{n+1}, \qquad \sum_{n=0}^{L-1} r_n^2=1
\end{equation}
which can be solved straightforwardly by numerical root-finding methods
(see figure \ref{fig1}). 
Eq.~\eqref{Hamiltonian} provides a relation between the shape of the breather and its energy, i.e., a relation between $\alpha$ and the energy
\begin{equation}\label{eq2.7}
E_\alpha=\sum_{n=0}^{L-1} \left( 2r_n^2-r_n r_{n-1}-r_n r_{n+1}-\frac{\alpha}{2} r_n^4\right)\, .
\end{equation}
For large $|\alpha|$, the $r_n$ and $\Omega$ can be determined by performing an expansion of all these quantities in powers of $1/\alpha$ and keeping the leading terms only. If the breather has its maximum at lattice site $n=c$, i.e.
$r_c \geq |r_n|$, one obtains 
\begin{eqnarray}\label{eq2.8}
r_c &=& 1-\frac 1 {\alpha^2} + \mathcal{O}(|\alpha|^{-3})\nonumber \\
r_{c\pm 1}&=& \frac 1 \alpha +\frac 1 {\alpha^3}+\mathcal{O}(|\alpha|^{-4}) \nonumber \\
r_{c\pm k}&=&\alpha^{-|k|} +\mathcal{O}(|\alpha|^{-(|k|+1)}) \nonumber \quad\hbox{ for } \quad k \ge 2\\
\Omega&=&-\alpha+2 + \mathcal{O}(|\alpha|^{-2})\nonumber\\
E_\alpha &=&-\frac{\alpha}{2}+2-\frac{2}{\alpha}+\mathcal{O}(|\alpha|^{-2}) \, .
\end{eqnarray}
The DNSE admits in the focusing case, i.e., for $\alpha>0$,
a breather, which is unimodal since all $r_n$ are positive. 
This solution can be mapped onto a solution for the defocusing case ($\alpha < 0$) by the transformation $r_n \to (-1)^{n}r_n$ (see section \ref{sec:2.1}), giving a staggered breather with the same unimodal envelope (see figure \ref{fig1}).

When $\alpha$ is very small and positive, the breather becomes very broad, and the difference $r_n-r_{n-1}$ becomes small. In such a continuum limit
the amplitudes $r_n$ can be captured by a smooth function $\Phi(x)$ via the scaling
$r_n=\Phi(n/L)/\sqrt{L}$. In such a continuum limit, we assume, for simplicity, a system of length one, in non-dimensional units. Then the normalisation condition in eq.~\eqref{eq2.6} becomes
the Riemann sum of the integral constraint $\int_0^1 |\Phi(x)|^2 dx=1$
and the discrete second derivative $2r_n-r_{n-1}-r_{n+1}$ to leading order can be written as 
$-\Phi''(x) / L^2$ with $x=n/L$. Hence, eq.~\eqref{eq2.6} becomes
\begin{equation}\label{eqDelta}
    0= - \Phi''(x) - \Omega L^2 \Phi(x) - \alpha L |\Phi(x)|^2 \Phi(x)\, .
\end{equation}
To obtain a well defined continuum limit, $L\rightarrow \infty$, we impose the
scaling relations $\Omega\sim L^{-2}$ and $\alpha\sim L^{-1}$, so that the profile
of the breather solution is captured by eq.~\eqref{eqDelta}. This limit permits a formal expansion in powers of $\alpha$, which is complementary to eq.~\eqref{eq2.8}, the asymptotic of the breather for large values of $\alpha$.

In contrast, for negative $\alpha$ such an approach cannot be used directly due to the staggered character of the breather, which means that the continuum limit
for $\alpha\rightarrow 0-$ is singular.

The stability of the breather states can be determined in terms of a linear stability analysis. Using the notation $c_n(t)=\exp(i\Omega t)(r_n + \delta r_n(t)+i\delta \varphi_n(t))$ in eq.~\eqref{eqofmotion} and expanding to linear order in the $\delta r_n$ and $\delta \varphi_n$, one obtains the  eigenvalue problem
\begin{align}
\label{eq2.9}
\Lambda
\left( \begin{array}{c}
\delta r_n \\ \delta \varphi_n
\end{array}\right)
&=\left(\begin{array}{cc}
0 &-(2-\Omega -\alpha r_n^2)\\
2-\Omega -3 \alpha r_n^2 & 0
\end{array}\right)
\left( \begin{array}{c}
\delta r_n \\ \delta \varphi_n
\end{array}\right)\nonumber\\
&+
\left(\begin{array}{cc}
0 & -1\\
1 & 0
\end{array}\right)
\left( \begin{array}{c}
\delta r_{n-1} \\ \delta \varphi_{n-1}
\end{array}\right)\nonumber\\
&+
\left(\begin{array}{cc}
0 & -1\\
1 & 0
\end{array}\right)
\left( \begin{array}{c}
\delta r_{n+1} \\ \delta \varphi_{n+1}
\end{array}\right) \, .    
\end{align}

\begin{figure}[!ht]
    \centering
   \includegraphics[width=0.8\linewidth]{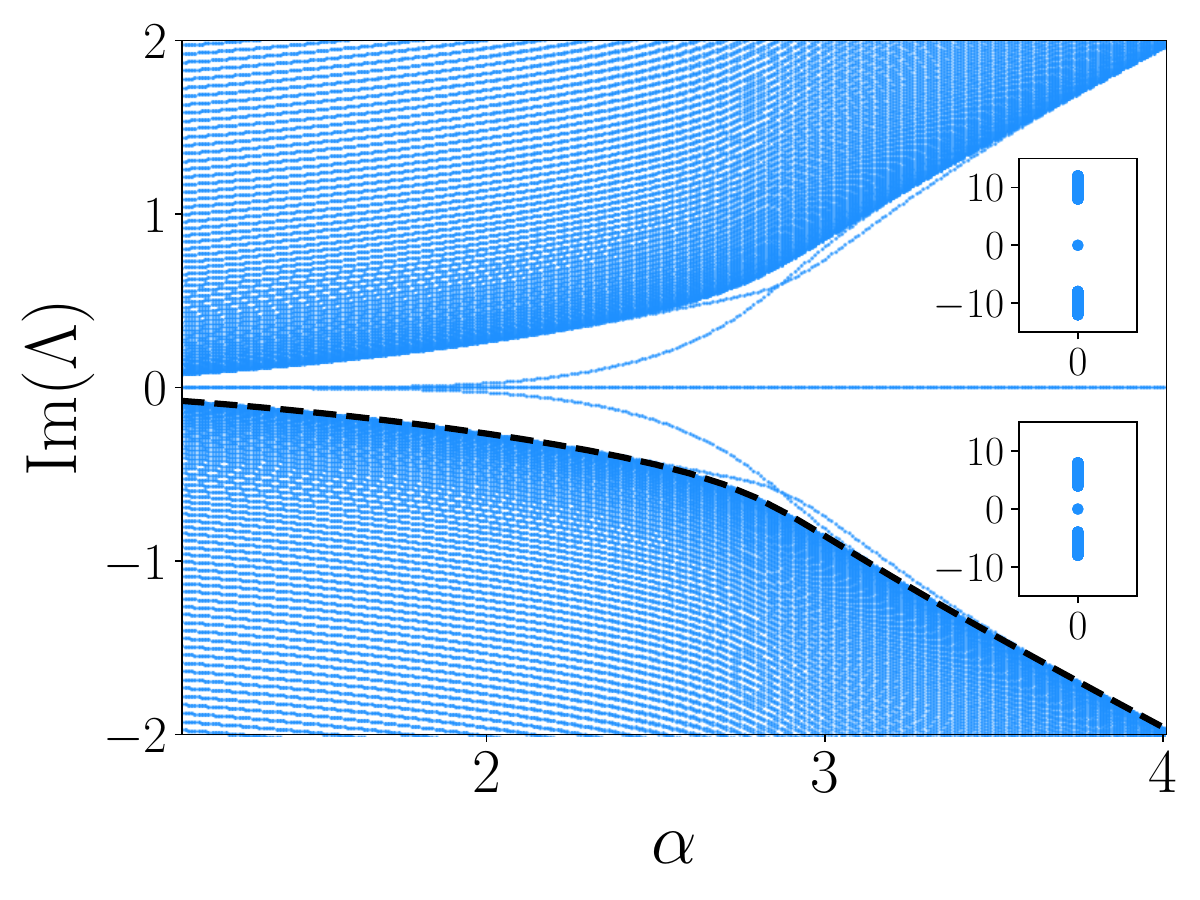}
    \caption{Eigenvalues (imaginary part) in dependence on $\alpha$ of the linear stability problem for unimodal breather solutions, see
    eq.~\eqref{eq2.9}, computed for a system of size $L=256$. The real part vanishes. The broken line shows the dependence of the breather frequency $\Omega$ on $\alpha$ (cf. Fig.~\ref{fig1}). The two insets
    show the location of eigenvalues in the complex plane for $\alpha=10$ (upper inset) and $\alpha=6$ (lower inset).}
    \label{fig2}
\end{figure}

Such an equation holds for every $n$, which means that the eigenvalues of a $2L \times 2L$ matrix must be evaluated for stability. This matrix has blocks of size $2\times 2$ on the diagonal and on the first off-diagonals. 
Because of the symplectic structure, eigenvalues come in quadruples 
$\pm \Lambda, \pm \bar{\Lambda}$. In our case, eigenvalues are always imaginary (see figure \ref{fig2}). While in general such a feature is not a sufficient condition for stability of a fixed point in Hamiltonian systems
(see e.g. \cite{Pfir_ZNA90}) in our case there is consensus that such
a breather corresponds to an extremum of the Hamiltonian and thus inherits stability (but not asymptotic stability) for any non-vanishing value of $\alpha$.
A breather solution where the maximum of the shape occurs on a bond and not at a lattice site yields eigenvalue pairs with finite real parts, and such breathers are linearly unstable.
All these features are well known (see e.g. \cite{EiJo:03}), but to the best of 
our knowledge
a formal analytic proof covering the whole range of $\alpha$ values has not been
established so far. 

The structure of the spectrum shown in Fig.~\ref{fig2} can be derived from eq.~\eqref{eq2.9} with the usual arguments borrowed from scattering theory.
First there occurs a pair of zero eigenvalue caused by the phase and the time translation invariance of the 
original equation of motion. This  pair of Goldstone modes is related to the conservation of $H$ and $N$, since conservation of $H$ leads to time translation invariance, and conservation of $N$ to invariance under a change of phase.
 The other eigenvalues are imaginary and are grouped in two
quasi-continuous bands with $\mbox{Im}(\Lambda)\in[|\Omega|,|\Omega|+4]
\cup[-|\Omega|-4|,-|\Omega|$ (see the two insets in figure \ref{fig2}) with the
corresponding eigenfunctions behaving asymptotically like plane waves. When $\alpha$
decreases these two bands approach each other and two isolated eigenvalues detach from these quasi-continuous bands, with the corresponding eigenfunctions being exponentially
localised. One pair of these localised eigenstates approaches zero at an exponential rate. This eigenstate is the precursor of another Goldstone mode which occurs in the continuum limit and which is related with the space translation invariance. 
The region shown in detail in figure \ref{fig2} shows a qualitative transition for $\alpha$ values at about 3.
 For large values of $\alpha$ the frequency $\Omega$ and the eigenvalues are to leading order linear functions of $\alpha$ with slope 1, see eq.(\ref{eq2.9}).
This behaviour corresponds to the so-called anti-integrable limit of the DNSE. This feature persists down to $\alpha$ values at around three, where a substantial curvature becomes noticeable and an expansion based on the anti-integrable limit may break down. The behaviour of the spectrum for smaller values of $\alpha$ seems to be dominated by the continuum limit $\alpha\rightarrow 0$, and the transition between these two regimes seems to be surprisingly sharp.

\subsection{Localised initial condition and time evolution}
If the DNSE is viewed as an effective equation of motion for a 
many-particle wave function, a state $c_n(0)=\delta_{n,c}$ that is fully localised at a lattice site $n=c$  is a natural initial condition. Such an initial state is close to a breather state, at least for larger $\alpha$,  and the question arises whether the wave function remains localised  under time evolution. 

Computation of solutions of eq.~(\ref{eqofmotion}) by numerical means
requires some care. Bog standard numerical integration schemes for ordinary differential equations normally do not preserve energy and thus induce uncontrollable drifts which render the long time results meaningless. Hence symplectic integrators are needed to deal with Hamiltonian dynamics \cite{HaLuWa:06}, so that
the energy is preserved at least approximately, which means that energy fluctuations stay bounded in time. Unless Hamiltonians have a very peculiar structure such symplectic
schemes may result in implicit integration schemes, and there is no a priori guarantee that
they preserve any other constant of motion, such as the particle number $N$.
Following for instance the basic exposition of
\cite{ShGeBoPaEg_PLA14} which was inspired by previous work \cite{Yosh_PLA90} on leapfrog integration methods, we use here an explicit symplectic
integration scheme which by design will also preserve $N$. The main idea is to factorise the time evolution operator $\exp(i\mathbf{L} \Delta t)$ by decomposing 
$\mathbf{L}$ into the linear nearest-neighbour coupling term and the nonlinear onsite term, both of which are integrable 
(see appendix \ref{sec:a.1} for more details). 

\begin{figure}[!ht]
    \centering
    \makebox[\linewidth]{{\includegraphics[width=6cm]{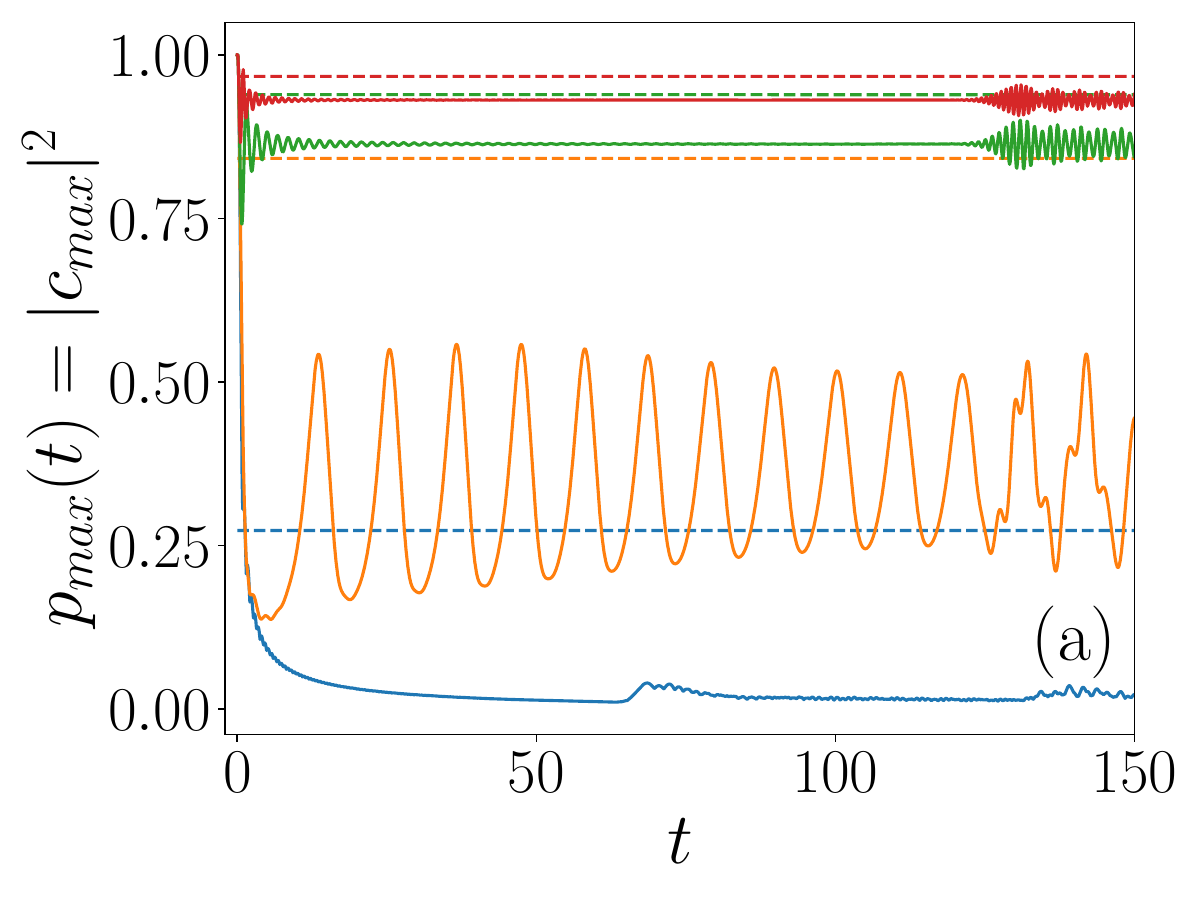} \hspace{0.1cm} {\includegraphics[width=6cm]{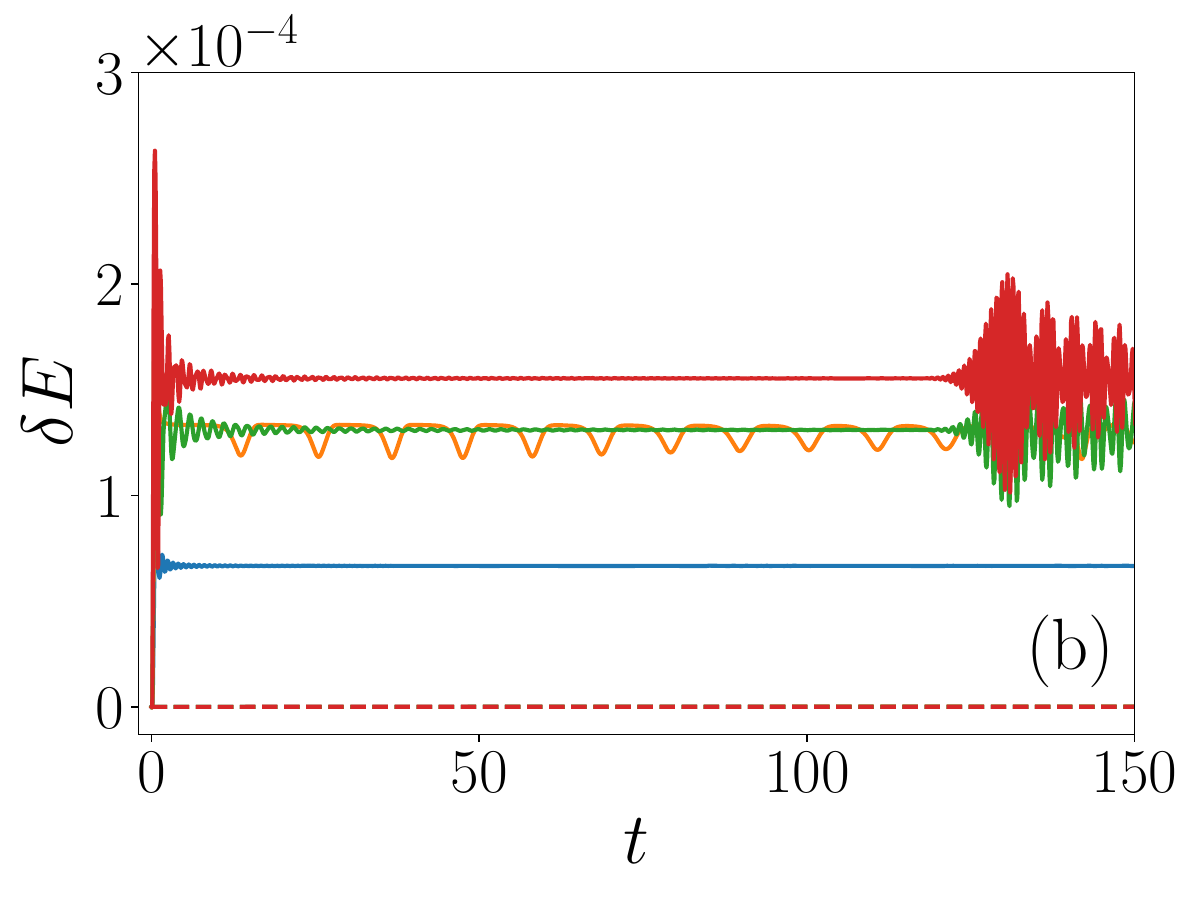} }}}
    \makebox[\linewidth]{{\includegraphics[width=6cm]{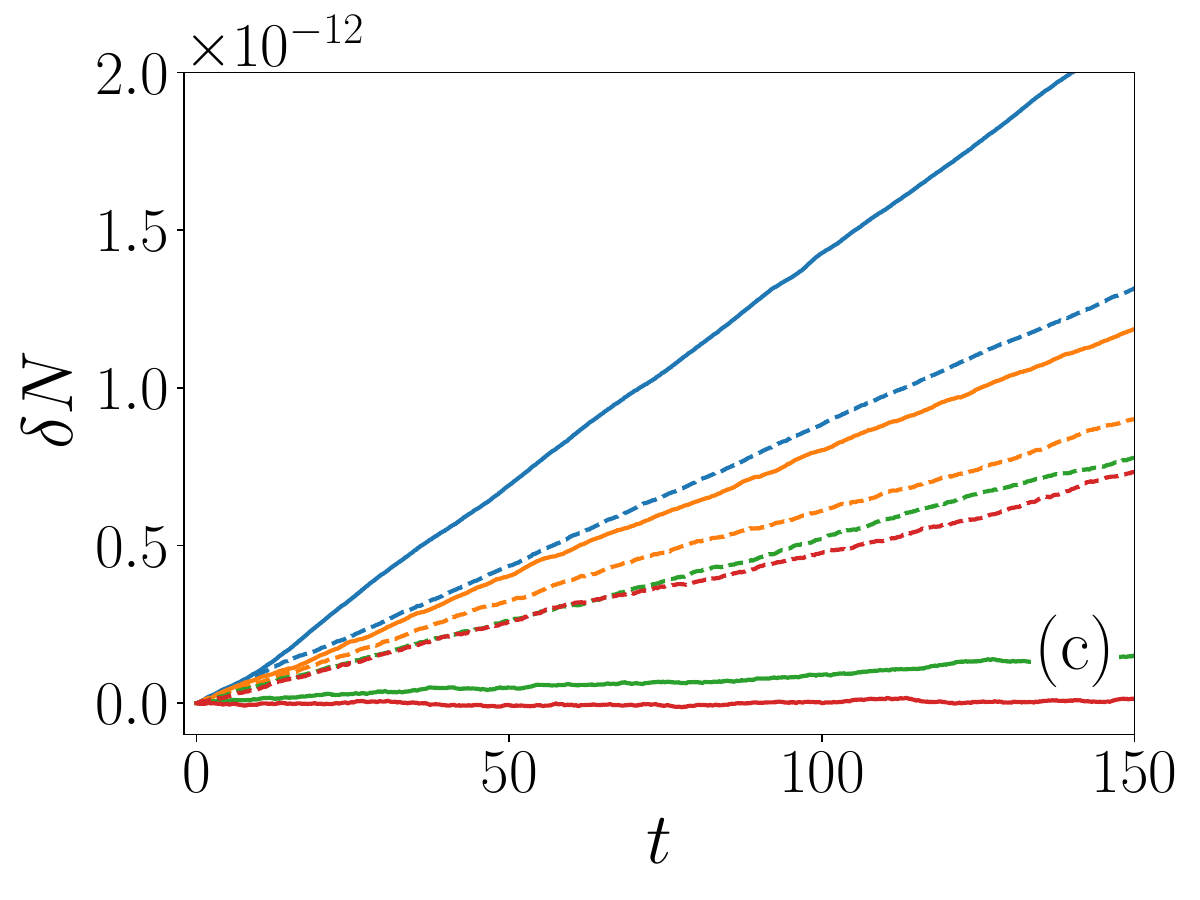} \hspace{0.1cm} {\includegraphics[width=6cm]{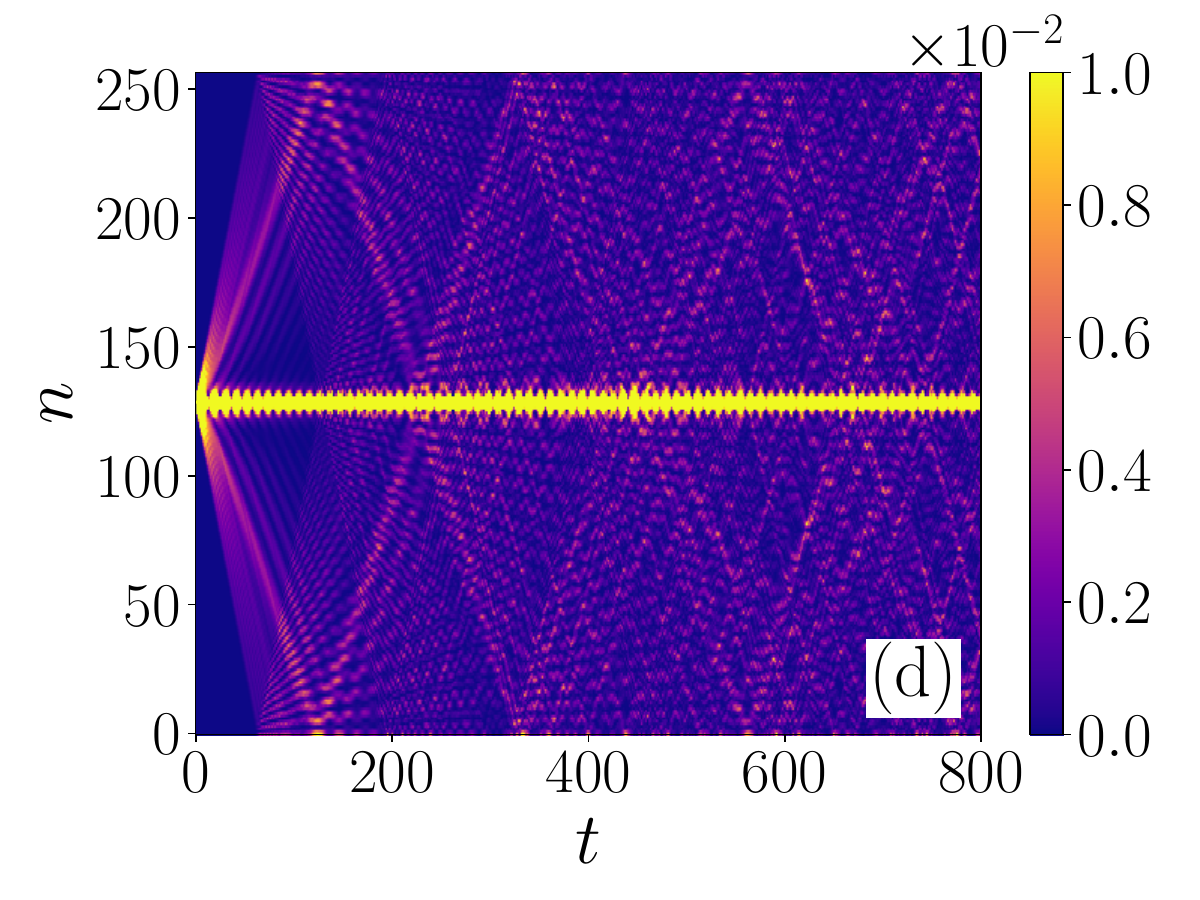} }}}
    \caption{Time evolution of eq.~\eqref{eqofmotion} for system size $L=256$ and with $\delta$-peak initial condition $c_n(0)=\delta_{n,128}$. The numerical integration was done using a symplectic integration scheme (see appendix \ref{sec:a.1} for details) with stepsize
    $\tau=0.01$. (a) Time traces of the amplitude $p_{max}(t)$ (the maximum of $|c_n(t)|^2$) for different values of $\alpha$: $\alpha=2$ (blue line), $\alpha=4$ (orange line),
    $\alpha=6$ (green line), and $\alpha=8$ (red line). The dotted horizontal line shows the maximal probability of the corresponding stable breather state. (b) Time evolution of the deviation of the energy from the initial energy $\delta E(t)=E(t)-E_{init}$ and
    (c) the deviation of the normalisation $\delta N(t)=N(t)-1$ for the same set of $\alpha$ values as in part (a). Full lines show the data when the $\delta$-peak initial condition has been used, broken lines show the data when the corresponding breather state has been used as initial condition. (d) Spatio temporal density plot for the evolution of $|c_{n}(t)|^2$ for $\alpha=4$ with emission of plane waves from the central peak and recurrence due to periodic boundary conditions, The peak exceeds 0.1, but we chose this colour scheme to highlight the recurring fluctuations in greater detail. (see as well panel (a) and (b)). }
    \label{fig3}
\end{figure}

Using an initial condition $c_n(0)=\delta_{n,c}$ we compute time traces of
the energy $E(t)$, the normalisation $N(t)$, and the largest
probability $p_{max}(t)=\max\{|c_n(t)|^2 : 1\leq n \leq L\}$ 
for different values of $\alpha$
(see figure \ref{fig3}). By design, the values
of energy and normalisation are preserved, and the small numerical fluctuations
give a rough impression of the numerical accuracy (see figure \ref{fig3}(b) and
\ref{fig3}(c)). The maximal
probability $p_{max}(t)$ is a simple proxy for the degree of localisation of the
state $\{c_n(t)\}$. If that value is close to 1, the state is almost
$\delta$ localised. Smaller time-dependent values that oscillate but do not decay 
still indicate some degree
of localisation. Decay towards very small values of $p_{max}(t)$ 
indicates that localisation is lost and the wave function becomes delocalised.

For large values of $\alpha$, the energy of the initial condition  $E_{init}=2-\alpha/2$ (see eq.~\eqref{Hamiltonian}) is almost identical to the energy of the breather state (see eq.~\eqref{eq2.8}) and the resulting time-dependent state performs small oscillations in the vicinity of the stable breather solution. With decreasing $\alpha$, the distance of the initial condition from the breather state increases so that the extent of localisation decreases, and the oscillations of the time-dependent solutions become more pronounced (see figure \ref{fig3}(a)). 
The localised state breaks down at about $\alpha \approx 3.5$ for $L=1024$. Such a breakdown is, however, not a signature of any dynamical instability. The particular initial condition is not any longer contained in the region of the phase space which
constitutes the elliptic island of the stable breather solution
(see as well, for instance, \cite{kevrekidis_discrete_2009} p.321).
The time-dependent dynamics of the localised solution is accompanied by a recurrence phenomenon whereby the localised part of the solution emits plane waves which recur after finite time, due to periodic boundary conditions (see figure \ref{fig3}(d)), causing an approach towards a stationary state where a localised peak oscillates in a chaotic fashion.

\section{The Stochastic Discrete Nonlinear Schr\"odinger Equation (SDNSE)}\label{sec:3}

We want to study the stability of the breathers with respect to white noise. We will here address two choices of stochastic perturbations, a multiplicative noise that preserves the normalisation, and a simple generic additive noise. To keep mathematical formalities at a minimum we will consider noise in the sense of Stratonovich. In physics terms that means the noise is supposed to have a very small but finite correlation time, which can be disregarded at the time scales of interest. In particular, one does not need to employ any sophisticated stochastic calculus and can treat the noise merely as an ordinary function when computing solutions of a stochastic differential equation. For this reason, we can again apply the symplectic integration scheme by extending it to non-autonomous systems and pooling the noise term together with the nonlinear term when factorising the time evolution operator (see appendix \ref{sec:a.2}). 

\subsection{Multiplicative Noise Model} \label{sec:3.1}

When a multiplicative noise term is included, the equation of motion (\ref{eqofmotion}) becomes
\begin{align}
    \label{eq3.1}
    \dot{c}_n(t) = &i \left( 2c_n(t) - c_{n-1}(t) - c_{n+1}(t) - \alpha |c_n(t)|^{2}c_n(t)\right) \nonumber\\
    + &i \sigma c_n(t) \xi_n(t) \, .
\end{align}
Here $\xi_n (t) \in \mathbb{R}$ denotes real-valued uncorrelated Gaussian random white noise with correlation function
\begin{equation}\label{eq3.2}
\langle \xi_n(t) \xi_m(s)\rangle = \delta_{n,m} \delta (t-s)\, ,
\end{equation}
and the parameter $\sigma \in \mathbb{R}$ determines the strength of the noise.
Eq.~\eqref{eq3.1}) constitutes still a Hamiltonian system with time-dependent
stochastic Hamiltonian
\begin{eqnarray}
    \label{eq3.3}
    H_\sigma(t) &=& \sum_{n=0}^{L-1} \Big(2|c_n|^2 - c_n \bar{c}_{n-1} - c_n \bar{c}_{n+1} 
    - \frac{\alpha}{2}|c_n|^4 \nonumber \\
    &&+ \sigma |c_n|^2 \xi_n(t)\Big) \nonumber \\
    &=& H + \sigma \sum_{n=0}^{L-1} |c_n|^2 \xi_n(t)
\end{eqnarray}

as can be easily confirmed using eq.~\eqref{eq2.2}. Since $\{H_\sigma(t),N\}=0$ we obtain conservation of the normalisation even in this stochastic case so that a multiplicative Stratonovich noise does not affect the normalisation. 

\begin{figure}[!ht]
    \centering
    \makebox[\linewidth]{{\includegraphics[width=7cm]{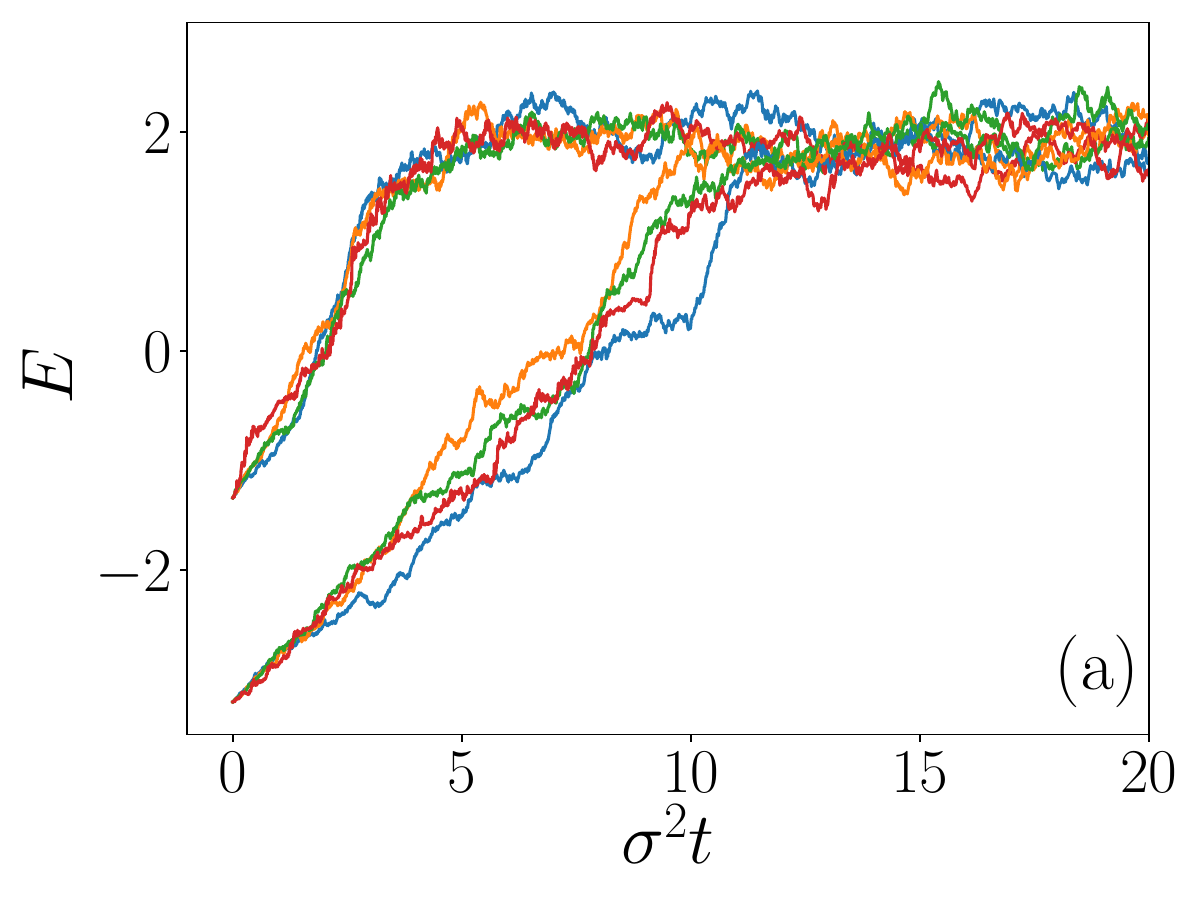} \hspace{0.1cm} {\includegraphics[width=7cm]{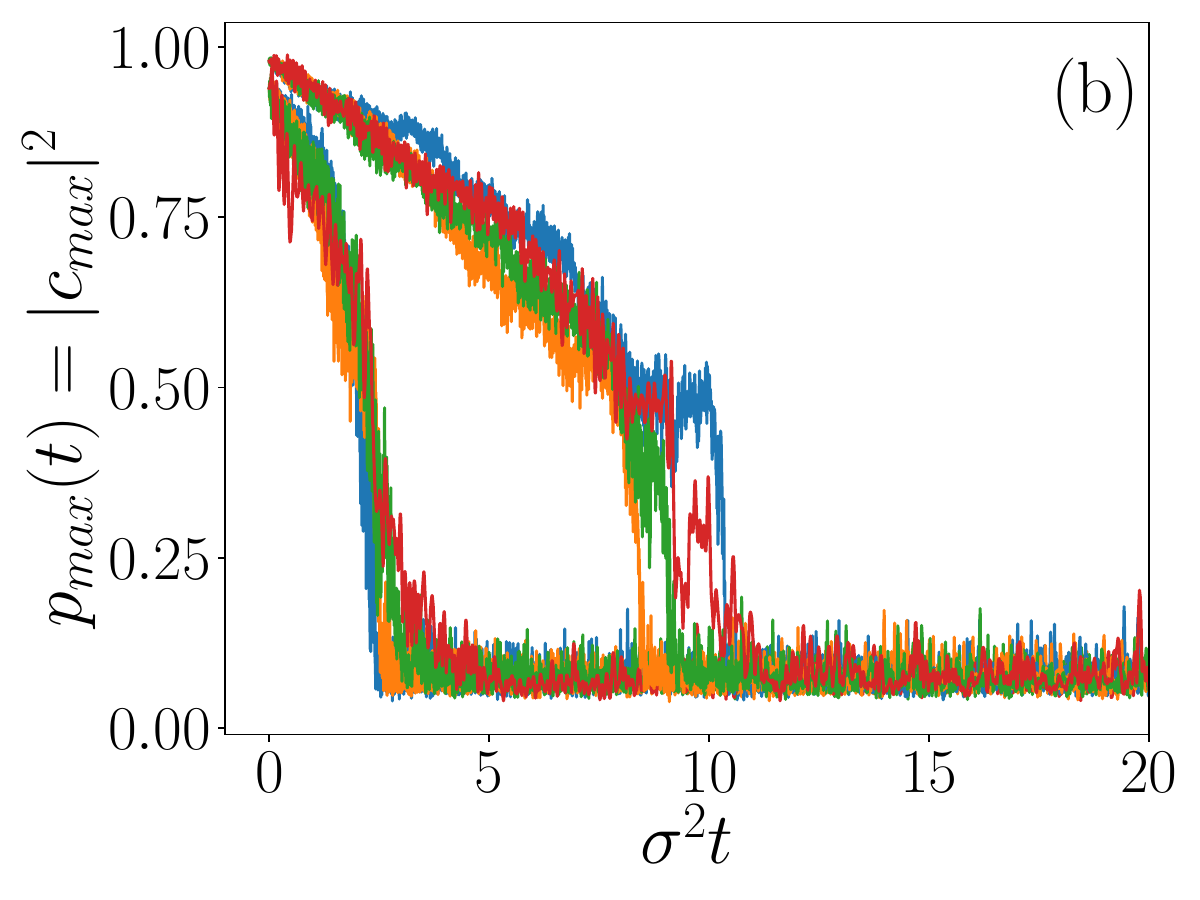}}}}
    \caption{Time traces obtained from numerical simulations of the multiplicative noise model eq.~\eqref{eq3.1}) for system size $L=64$ and different noise levels: $\sigma = 0.01$ (blue), $\sigma=0.03$ (orange), $\sigma=0.3$ (green), and $\sigma=0.1$ (red). Data are shown on a rescaled time scale $\sigma^2 t$. (a) Time dependence of the energy, eq.~\eqref{Hamiltonian}, for $\alpha=6$ (upper set of curves) and $\alpha=10$ (lower set of curves). (b) Time dependence of the largest probability $p_{max}(t)$ for $\alpha=6$ (lower set of curves) and $\alpha=10$ (upper set of curves). For all $\alpha$ the corresponding deterministic breather state was chosen as initial condition (cf. eq.~\eqref{eq2.8}).}
    \label{fig4}
\end{figure}

To study the robustness of the breather solution against multiplicative noise, we perform
numerical simulations of the stochastic differential equation \eqref{eq3.1} for
various parameter values, where we take the particular breather as an initial condition.
In the literature, one can find sophisticated numerical integration schemes for symplectic stochastic systems (see e.g.~\cite{HaLuWa:06} for a comprehensive review) which are implicit and thus more difficult to implement, and there is a priori no guarantee that such schemes preserve a constant of motion.  In contrast, our explicit higher-order integration scheme (see appendix \ref{sec:a.2}) observes the conservation of $N$  by design and is straightforward to apply. The resulting time traces for the energy and the maximal probability are shown in figure \ref{fig4} for a range of noise amplitudes $\sigma$ and $\alpha$ values that support a localised breather state. The energy seems to depend on the effective time variable $\sigma^2 t$ only, in line with simple diffusion processes. The energy increases initially linearly, with a slope roughly proportional
to $\alpha^{-1}$, and then saturates with fluctuations around $E\approx 2$. 
The maximal
probability decays as well on a joint time scale $\sigma^2 t$. The localised state is destroyed by the noise in favour of a delocalised state, which spreads across the system.

\begin{figure}[!ht]
\includegraphics[width=0.48 \textwidth]{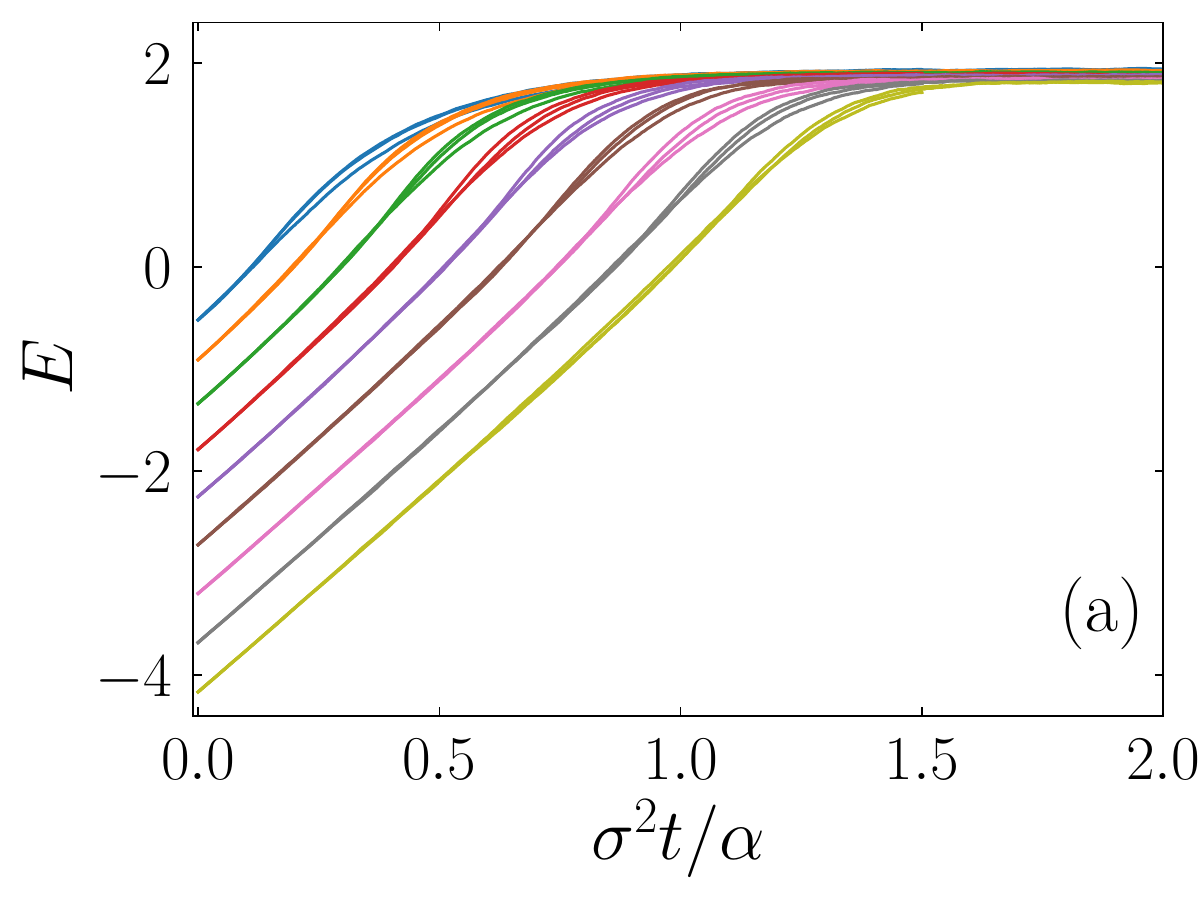}
\includegraphics[width=0.48 \textwidth]{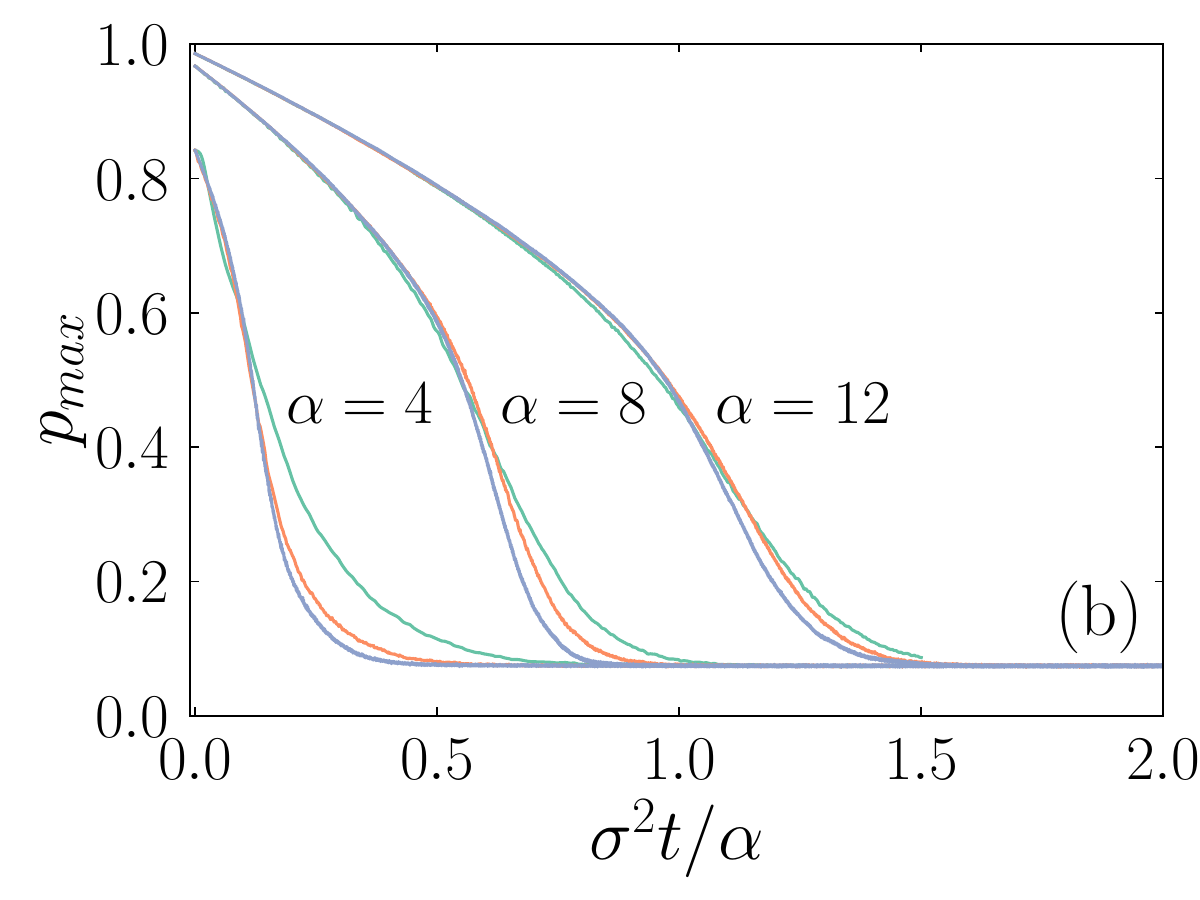}
\includegraphics[width=0.48 \textwidth]{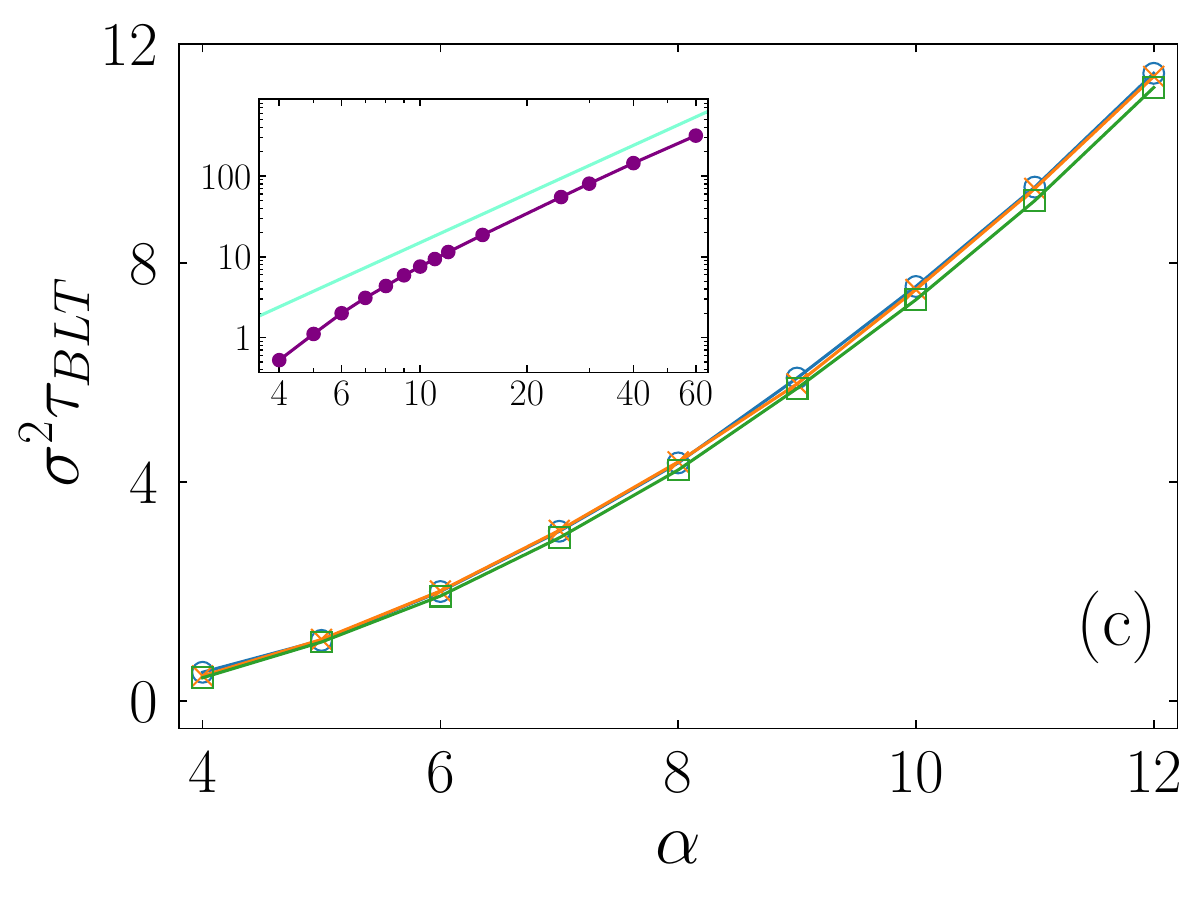}
\includegraphics[width=0.48 \textwidth]{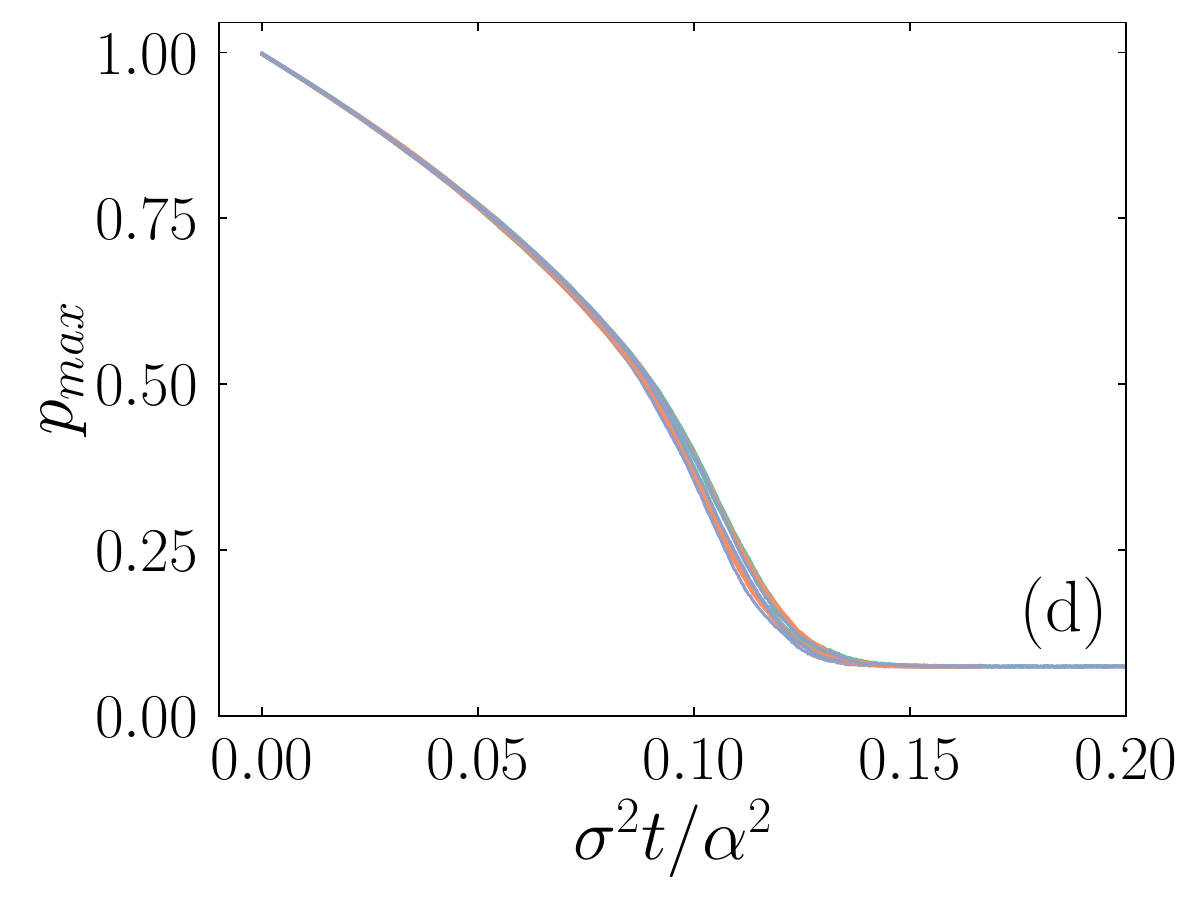}
\caption{Ensemble averages of time traces of
the SDNSE, eq.~\eqref{eq3.1}, with system size $L=64$ and with breather state as initial
condition (cf. eq.~\eqref{eq2.8}). Means have been computed over $10^3$ realisations of
the noise.
(a) Mean energy $E$ as a function of a rescaled time, $\sigma^2 t/\alpha$, for different values of $\alpha$ (from left to right: $\alpha=4$, $5$, $6$, $\ldots$, 
$11$, $12$) and three values of the noise amplitude, $\sigma=0.3$, $0.1$ and $0.03$.
For given value of $\alpha$ the three curves for different noise amplitude almost overlap
at the scale of the graph. (b) Amplitude $p_{max}$ of the spatial pattern as a function
of the rescaled time, $\sigma^2 t/\alpha$, for different $\alpha$ and $\sigma$ values (green: $\sigma=0.03$,
orange: $\sigma=0.1$, blue: $\sigma=0.3$).
(c) Mean lifetime of the breather, $\tau_{BLT}$, in dependence of $\alpha$ for three different
values of the noise amplitude (square: $\sigma=0.03$, circle: $\sigma=0.3$, cross: $\sigma=0.1$, the lines are a guide to the eye). The inset shows the mean lifetime
on a double logarithmic scale (for $\sigma=0.3$). The straight line in the inset has slope $2$. (d) $p_{max}$ in dependence on the rescaled time variable $\sigma^2 t/\alpha^2$ for $\alpha = 25$, $30$, $40$ and $\sigma = 0.3$, $0.1$, $0.03$. All twelve curves overlap at the scale of the graph.}
\label{fig5_new}
\end{figure}

We can obtain a better quantitative analysis by looking at mean values averaged over
an ensemble of orbits. In all cases we take again the corresponding
breather state as initial condition. For the time dependence of the
energy we observe an almost perfect data collapse when a rescaled time variable
$\sigma^2 t/\alpha$ is employed, see figure \ref{fig5_new}(a). The mean energy
shows initially a linear increase with constant slope, where the observed offset for
different values of $\alpha$ is simply caused by the dependence of the initial energy
on $\alpha$. The finding that the initial slope of $\dot E $ is proportional to $1/\alpha$ can be explained to some extent by considering the parameter-free version eqs.~(6-8) of the model: If the shape of the decaying breather is essentially determined by its energy $\tilde H$, the time evolution of $\tilde H$ depends mainly on $\tilde H$ itself, irrespective of the value of $\alpha$. Using the relation $E \propto \tilde H/\alpha$, we find $\dot E \propto 1/\alpha$. 

The time evolution of the amplitude of the state, $p_{max}=\max\{|c_k|^2\}$ shows a
different characteristic, see figure \ref{fig5_new}(b).
As already reported in \cite{iubini_2017} there occur two different temporal regimes.
For smaller times, roughly speaking as long as $p_{max}(t)>1/2$, we obtain a quite
accurate data collapse with respect to the noise amplitude, so that the amplitude is
effectively a function of $\sigma^2 t$. For larger times, say $p_{max}(t) < 1/2$, that scaling
deteriorates and a mild additional dependence on the noise amplitude becomes
visible. Overall the distinction between these two temporal regimes
becomes more pronounced if the on-site potential strength $\alpha$ increases.
Furthermore,  the overall dependence of $p_{max}$ on $\alpha$ differs from the scaling
properties of the energy. To highlight this point further it is sensible to 
look a the lifetime of the breather state. There is no unique way to
assign such a lifetime. We adopt here the definition of a
(stochastic) lifetime such that
the value of $p_{max}(t)$ drops for the first time below half of its initial value.
The mean of this lifetime, $\tau_{BLT}$,  in dependence on $\alpha$ is shown in
figure \ref{fig5_new}(c).  We still observe the aforementioned scaling
of the lifetime with $\sigma^2$. The dependence of $\tau_{BLF}$ on $\alpha$ shows also
a quadratic behaviour for larger values of $\alpha$ (see as well the inset in
figure \ref{fig5_new}(c)). As a consequence, we observe a reasonable data collapse for larger values of $
\alpha$ when $p_{max}$ is considered in dependence of the rescaled time variable $\sigma^2 t/\alpha^2$, see 
figure \ref{fig5_new}(d). In particular, this behaviour shows that other definitions of a mean lifetime would give quantitatively analogous results. In summary, mean values 
seem to depend on the rescaled time $\sigma^2 t$ in the range of noise amplitudes employed here,
while the dependence on the on-site potential strength $\alpha$ is more subtle and
depends on the quantity under consideration. That the decay time of $p_{max}$ is for large $\alpha$ proportional to $\alpha^2$ can be made plausible by considering eqs.~(1) and (11): the time evolution of $p_{max}$ has a nonlinear term proportional to $\alpha$, which tends to preserve the maximum value, and the term that transfers weight to the two neighbors is proportional to $r_{c\pm 1}$, which in turn is proportional to $1/\alpha$. This is weaker by a factor $1/\alpha^2$ compared to the main term that counteracts destruction of the breather. 

The time traces suggest that the stochastic dynamics tends towards a stationary state that is characterised by a finite value of energy $E \approx 2$, irrespective of the value of $\alpha$ and of the initial condition. Indeed, it is plausible to assume that the system is ergodic: the noise term gives the $c_n$ random kicks that change their phases, and the nearest-neighbour coupling subsequently changes the amplitudes, while keeping the overall normalisation at one, so that the energy cannot escape to infinity. In this way, the noise can push the state of the system through phase space, as there is no restoring force that tends back to the initial state. Eventually, the boundary of the region of linear stability of the breather will be passed, and the long-term behaviour will be dominated by the maximum entropy distribution of the $c_n$. Ergodicity implies a long-term behaviour that results in a uniform distribution
of the $c_n$ on the surface of the $2L$-dimensional unit sphere 
\begin{equation}\label{eq3.4}
\rho_{mc}(\{c_n,\bar{c}_n\}) = \frac{1}{Z_{mc}}\delta\left(N-1\right)
\end{equation}
with
\begin{align}\label{eq3.5}
Z_{mc}&=\int_{\mathbb{C}^L} d(c_1,\bar{c}_1,c_2,\bar{c}_2,\ldots, c_L \bar{c}_L) \delta\left(N-1\right)\nonumber\\
&= \frac{\pi^L}{(L-1)!}
\end{align}
With a slight abuse of the standard notation
(see e.g. \cite{PhysRevLett.112.020602})
we call eq.~\eqref{eq3.4} simply a
"microcanonical" distribution, since $N$ is the conserved quantity in the
stochastic setup.

Our ergodic system of $L$ variables $c_n$ satisfies the constraint $\sum_n (\Re(c_n)^2 + \Im(c_n)^2) = N$, a condition which is formally equivalent to a system of $L$ independent harmonic oscillators for which the total energy $H=\sum_n (p_n^2 + q_n^2)$ is conserved. In this setup the equivalence of ensembles in the thermodynamic limit can be shown by elementary methods. The equivalent of a canonical ensemble for the harmonic oscillators (where the mean energy is determined by the temperature) is a grand canonical ensemble of our system, where a chemical potential $\nu$ determines the average particle number $\langle N \rangle$. Hence for large system size $L$ the "microcanonical" ensemble, eq.~\eqref{eq3.4}) is equivalent to the grand canonical ensemble 
\begin{equation}\label{eq3.6}
\rho_{gc}(\{c_n,\bar{c}_n\})=\frac{1}{Z_{gc}} \exp\left(-\nu N\right)
\end{equation}
with
\begin{equation}\label{eq3.7}
Z_{gc}=\frac{\pi^L}{\nu^L} .
\end{equation}
The chemical potential $\nu$ is calculated via
\begin{equation}\langle N \rangle_{gc} = -\frac{\partial}{\partial \nu} \ln Z_{gc} = \frac L \nu = 1\, ,
\end{equation}
giving $\nu = L$. 
Since the distribution eq.~\eqref{eq3.6} does not involve the Hamiltonian $H$, it corresponds formally to a situation of infinite temperature, which is also evident from the fact that the different $c_n$ are uncoupled in eqs.~\eqref{eq3.4} and \eqref{eq3.6}. 

\begin{figure}[!ht]
    \centering
    \makebox[\linewidth]{{\includegraphics[width=6.5cm]{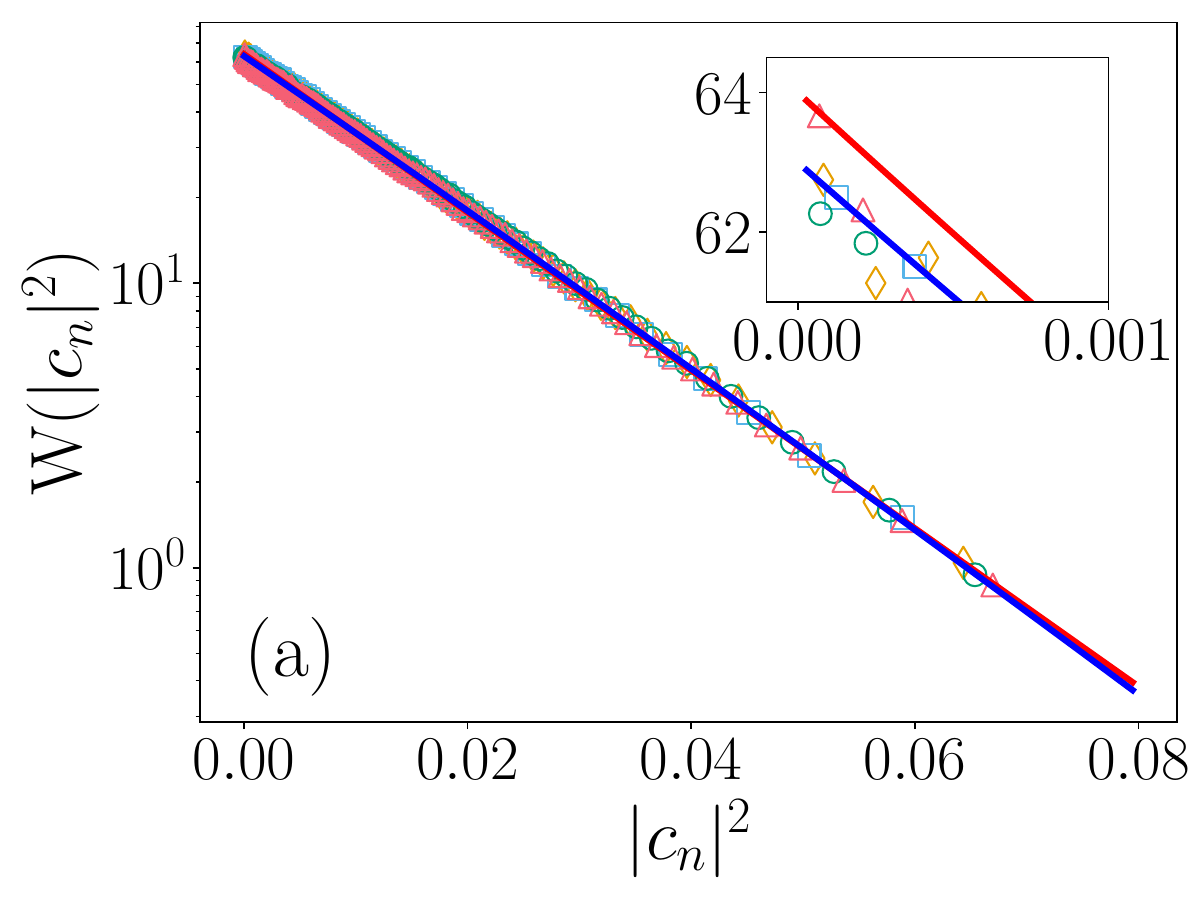} \hspace{0.1cm} {\includegraphics[width=6.5cm]{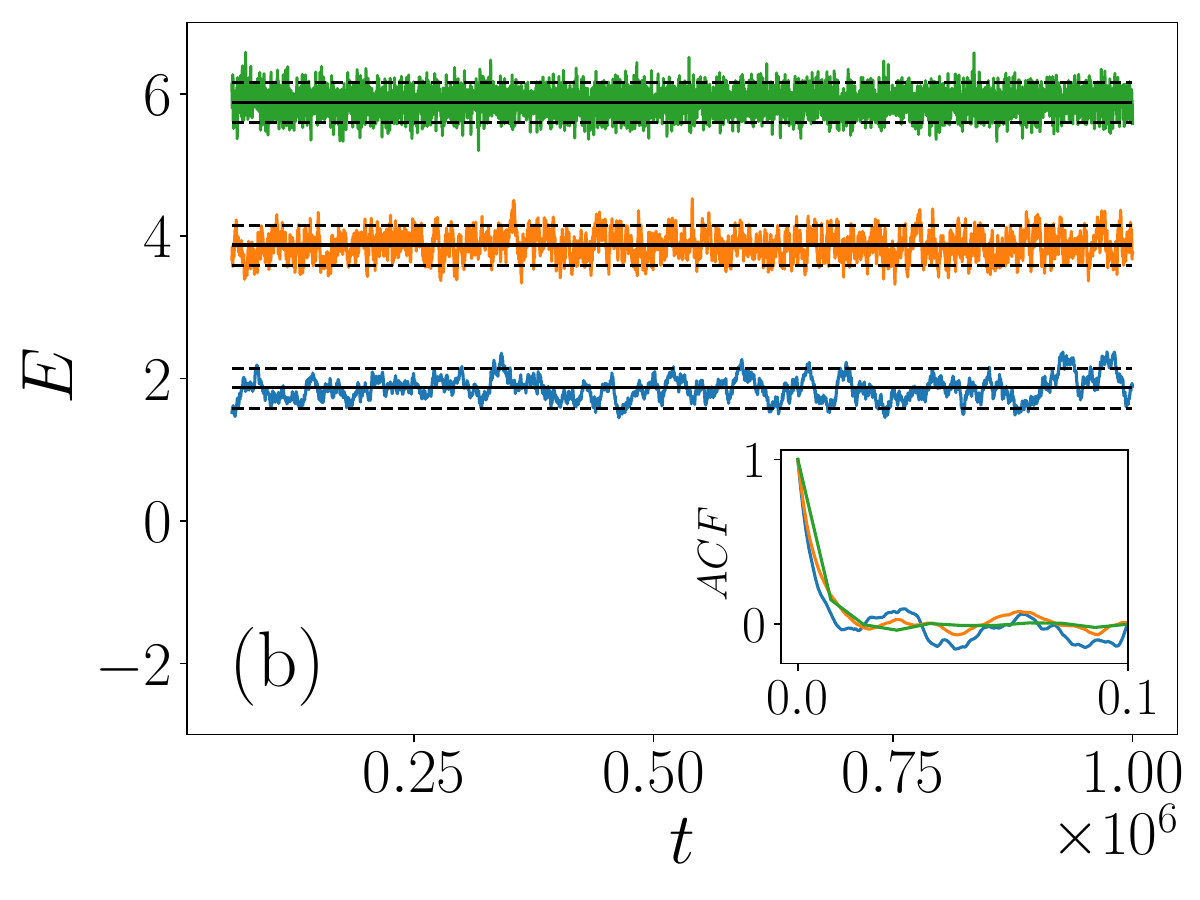}}}}
    \caption{(a) Marginal distribution $W(|c_n|^2)=\pi p(c_n,\bar{c_n})$ in the stationary state for the multiplicative noise model, eq.~\eqref{eq3.1}), for a system of size $L = 64$, $\sigma=0.1$ and different $\alpha$ values. Symbols: Histogram obtained from about $3.2\times 10^7$ data points of the stationary part of the time series and    
    $\alpha = 2$ (yellow,  diamond), $\alpha=4$ (blue, squares), $\alpha=6$ (green circles), and $\alpha=8$ (red triangles). Data are shown on a semi-logarithmic scale. Lines represent the analytic estimates obtained from the microcanonical (orange) and canonical (blue) ensemble, respectively (see eqs.~(\ref{eq3.8}) and (\ref{eq3.9})). The inset shows the data close to zero on a smaller scale. (b) Time traces of the energy, eq.~\eqref{Hamiltonian}, in the stationary state for $\alpha = 8$, a system of size $L=64$ and different noise levels: $\sigma = 0.01$ (blue), $\sigma=0.03$ (orange), and $\sigma=0.1$ (green), cf. Fig. \ref{fig4}(a). The different curves have been displayed with a mutual offset of $2$ for visibility. The solid and dashed black horizontal lines indicate the canonical estimate of the mean and the standard deviation of the energy in the stationary state (see eqs.~(\ref{eq3.10}) and (\ref{eq3.10a}), respectively). The inset shows the corresponding normalised autocorrelation function of the energy on a rescaled time scale $\sigma^2 t$. }
    \label{fig5}
\end{figure}

For comparison of these analytical predictions with the simulation data we resort to the marginal distributions, e.g., to the distribution of a single amplitude $c_1$. The microcanonical and the grand canonical distribution yield
\begin{eqnarray}\label{eq3.8}
p_{mc}(c_1,\bar{c}_1)&=&\int_{\mathbb{C}^{L-1}} d(c_2,\bar{c}_2,c_3,\bar{c}_3,\ldots, c_L \bar{c}_L) \rho_{mc}(\{c_k,\bar{c}_k\}) \nonumber \\
&=& \frac{L-1}{\pi} \left(1-|c_1|^2\right)^{L-2}, \quad (|c_1|^2\leq 1)
\end{eqnarray}
and
\begin{equation}\label{eq3.9}
p_{gc}(c_1,\bar{c}_1)=\frac{L}{\pi} \exp\left(-L |c_1|^2\right)
\end{equation}
respectively.
Both expressions, eqs.~\eqref{eq3.8} and \eqref{eq3.9}, virtually do not differ for moderate or large system size, and both coincide
with the numerical data obtained in the stationary state (see figure \ref{fig5}(a)).
However, small finite size corrections are discernible in the data close to zero (inset in figure \ref{fig5}(a)), which confirms that the actual state is captured by a
microcanonical distribution, eq.~\eqref{eq3.4}, as expected.
In addition, we can also look at the expectation of the energy to consolidate the analytic estimates. While the grand canonical distribution, eq.~\eqref{eq3.6}, describes independent random variables, the microcanonical ensemble, 
eq.~\eqref{eq3.4}, contains non extensive correlations. Nevertheless, 
we have $\langle c_n \bar{c}_{n-1}\rangle_{mc}=\langle c_n \bar{c}_{n-1}\rangle_{gc}=0$, because of phase symmetry. 
Furthermore, eqs.~\eqref{eq3.8} and \eqref{eq3.9} result in
$\langle |c_n|^2 \rangle_{mc}=\langle |c_n|^2 \rangle_{gc}=1/L$ and 
$\langle |c_n|^4 \rangle_{mc}=2/(L(L+1))$ as well as $\langle |c_n|^4 \rangle_{gc}=2/L^2$, i.e., there are finite size corrections visible when comparing the values of the fourth moment. 
Hence, the mean value of the energy eq.~\eqref{Hamiltonian} in the stationary state reads
\begin{equation}\label{eq3.10}
\langle H \rangle_{mc}=2-\frac{\alpha}{L+1}, \qquad
\langle H \rangle_{gc}=2-\frac{\alpha}{L} \, .
\end{equation}
These two values agree pretty well with the data obtained from direct numerical simulations
(see figure \ref{fig5}(b)), while the data do not allow us to decide which of the two values fits better because of the persistent fluctuations in the stationary state. In fact, the size of these fluctuations can be evaluated in terms of the
variance of the energy. If we use for simplicity grand canonical expectations,
i.e. $\langle |c_n|^{2k} |c_m|^{2\ell}\rangle_{gc}=(k!/L^k) (\ell!/L^\ell)$
for $n\neq m$, 
we have
\begin{equation}\label{eq3.10a}
\langle H^2 \rangle_{gc}-\langle H\rangle_{gc}^2
=\frac{6}{L}-\frac{8\alpha}{L^2}+\frac{5\alpha^2}{L^3} \, .
\end{equation}
The corresponding standard deviation fits the size of the energy fluctuations in the
stationary state perfectly well (see figure \ref{fig5}(b)). In particular, the size of the energy fluctuations does not depend on the noise strength as expected in a
microcanonical or grand canonical ensemble. The considerations so far are not able to clarify the dynamics in the stationary state. Time traces of the
energy show that correlations increase when the noise amplitude decreases (see 
figure \ref{fig5}(b)) so that, finally, the deterministic dynamics is restored. 
The computation of the autocorrelation function
of the energy is consistent with a correlation time that scales as
$T_{corr}\sim\sigma^{-2}$ (see inset in figure \ref{fig5}(b)), as one would
expect from phase diffusion and simple stochastic models like the Kubo oscillator \cite{Kubo:62}.

In summary, the numerical results confirm the expected behaviour that multiplicative noise heats up the system by entropy production. Thanks to the conservation of normalisation, finally, a stationary state emerges, characterised by an infinite temperature. 

\subsection{Additive Noise model} \label{sec:3.2}

As a second example for studying the robustness of breather states against noise, we consider the seemingly simpler case of additive noise, where the equation of motion reads
\begin{align}
    \label{eq3.11}
    \dot{c}_n(t) &= i \left( 2c_n(t) - c_{n-1}(t) - c_{n+1}(t) - \alpha |c_n(t)|^{2}c_n(t)\right)\nonumber\\
    &+ i \sigma \xi_n(t) \, .
\end{align}
Here, $\xi_n(t)$ denotes again uncorrelated real valued Gaussian noise with correlation given by eq.~\eqref{eq3.2}, and the real-valued parameter $\sigma$ denotes the strength of the noise. This system is, again, a stochastic Hamiltonian system with Hamiltonian
\begin{eqnarray}
    \label{eq3.12}
    H_\sigma(t) &=& \sum_{n=0}^{L-1} \Big(2|c_n|^2 - c_n \bar{c}_{n-1} - c_n \bar{c}_{n+1} -
    \frac{\alpha}{2}|c_n|^4 \nonumber\\
    &&+ \sigma (c_n+\bar{c}_n) \xi_n(t)\Big)\nonumber \\
    &=& H + \sigma \sum_{n=0}^{L-1} (c_n +\bar{c}_n) \xi_n(t) \, .
\end{eqnarray}
In this case, normalisation is no longer preserved since $\{H_\sigma(t),N\}\neq 0$, and we expect a completely different dynamical behaviour. For the numerical treatment, we again employ a simple higher-order stochastic symplectic integration scheme (see appendix \ref{sec:a.2} for some details).

\begin{figure}[!ht]
    \centering
    \makebox[\linewidth]{{\includegraphics[width=6.5cm]{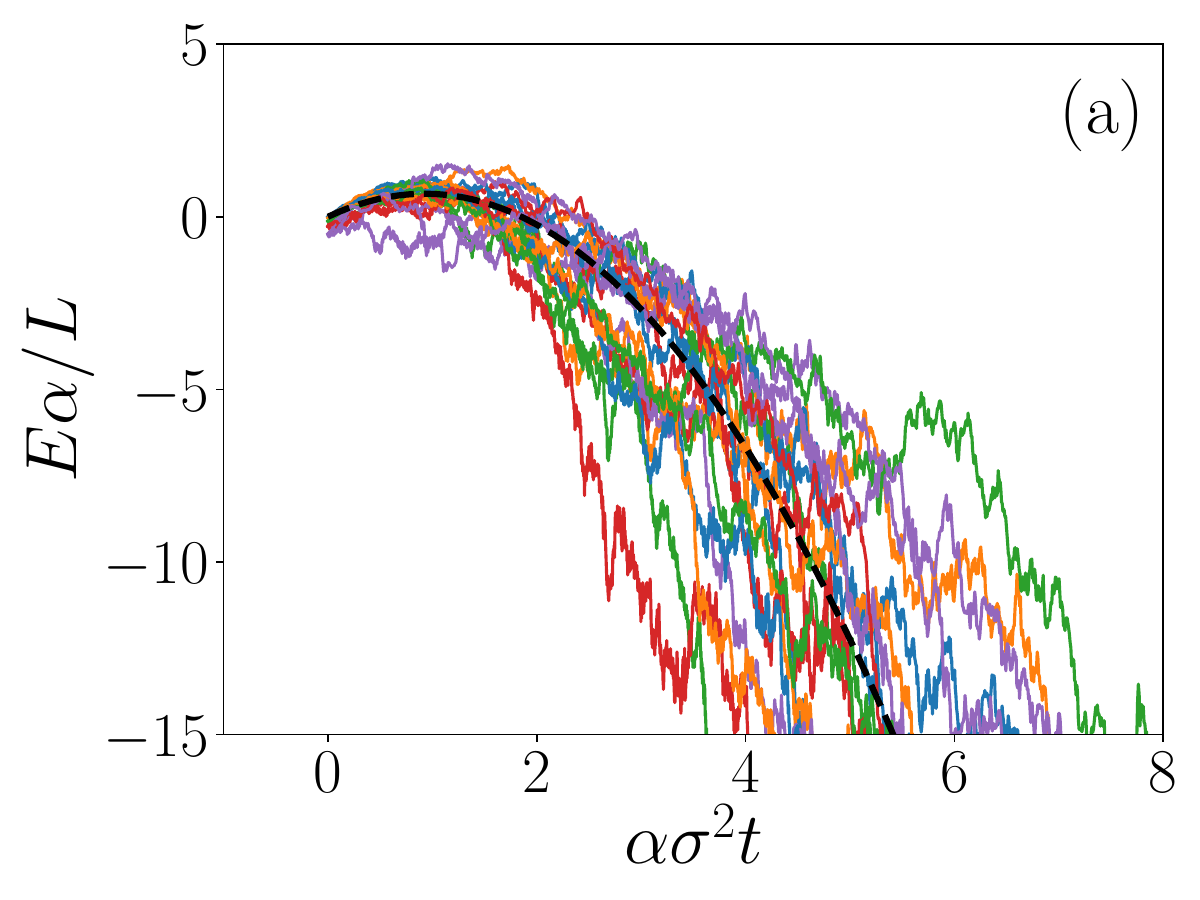} \hspace{0.1cm} {\includegraphics[width=6.5cm]{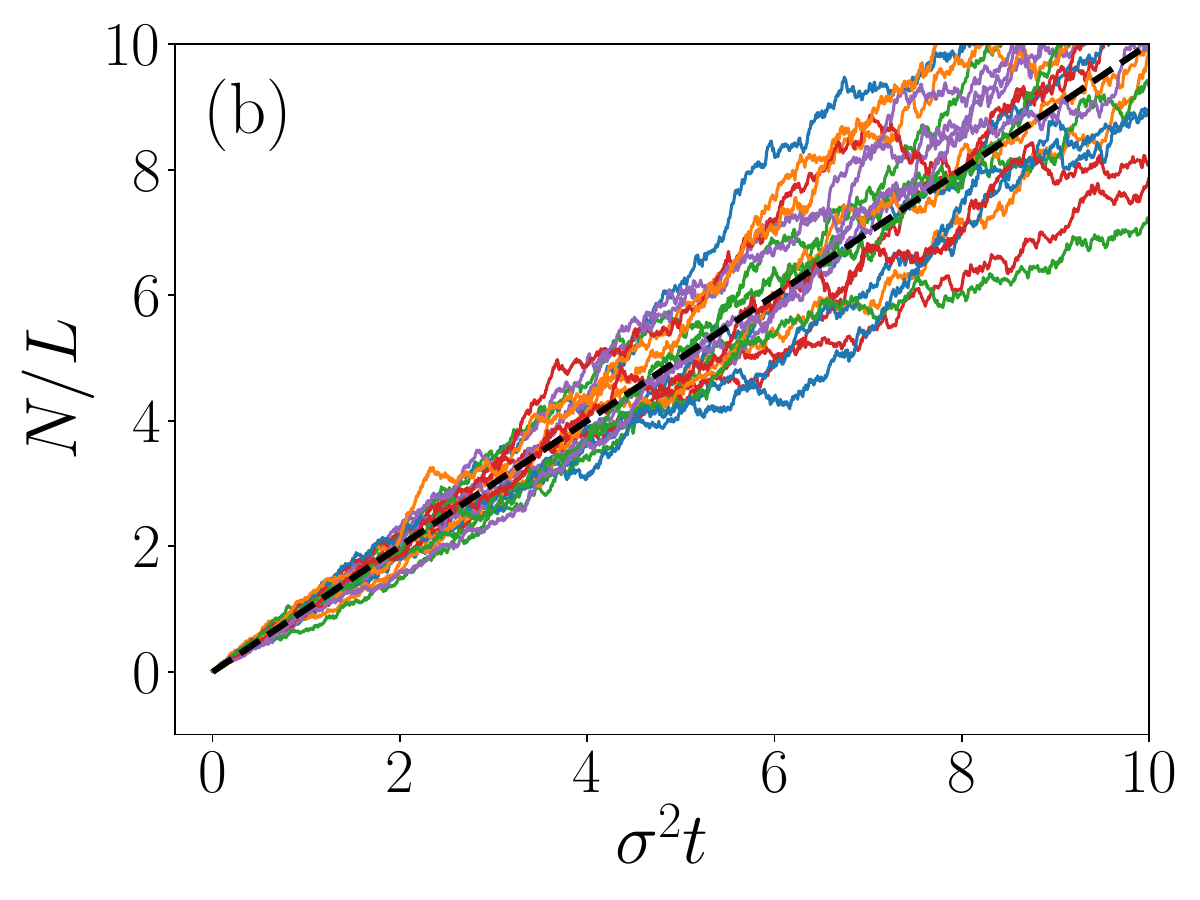}}}}
    \caption{Time traces of (a) the energy scaled by $\alpha$ and of (b) the normalisation obtained from numerical simulations of the additive noise model eq.~\eqref{eq3.12}) with system size $L=64$ and the breather state (cf. eq.~\eqref{eq2.8}) as an initial condition. Results are shown on a rescaled time scale, $\alpha \sigma^2 t$ and $\sigma^2 t$, respectively. Time traces have been computed for $\alpha = 2$ (blue), $\alpha=4$ (orange), 
    $\alpha=6$ (green), $\alpha=8$ (red), and $\alpha=10$ (purple) and a range of noise levels  $\sigma \in \{0.01, 0.03, 0.1, 0.3\}$. The dashed black lines are the simple analytic estimates $\alpha E/L=2 \alpha \sigma^2 t- 3 (\alpha \sigma^2 t)^2/2$ and $N/L=\sigma^2 t$, respectively (see the text for details).
    }
    \label{fig6}
\end{figure}

Using again the corresponding breather as a localised initial condition,
the time traces show an almost linear increase of the normalisation $N(t)$ and an unbounded energy $E(t)$ for a large range of parameter values (see figure \ref{fig6}). In fact, when plotting data on a suitably rescaled time scale, one obtains a reasonable, albeit not perfect, data collapse. One can give a simple heuristic explanation for these numerical
findings. While the deterministic part of eq.~\eqref{eq3.11} preserves the normalisation, the additive stochastic force changes $N$. Hence, we may assume
that the stochastic contribution becomes dominant, and the very coarse approximation
$\dot{c}_n(t) \sim i \sigma \xi_n(t)$ captures the essence of the phenomena. Hence,
the amplitudes $c_n$ obey Brownian motion, and the normalisation follows the
average path $N(t) \sim L \sigma^2 t$. This simple reasoning fits the data quite
well over a large range of parameter values (see figure \ref{fig6}(b)). If we use the same reasoning to evaluate the energy, eq.~\eqref{Hamiltonian}, we obtain the estimate $E\sim 2 L \sigma^2 t-3 L \alpha (\sigma^2 t)^2/2$ so that 
$\alpha E\sim L(2 \alpha \sigma^2 t-3 (\alpha \sigma^2 t)^2/2)$. Again such a simple expression fits the data quite well (see figure \ref{fig6}(a)), even though we do not obtain a perfect data collapse.
Hence, to leading order the dynamics of the additive noise model, eq.~\eqref{eq3.11}, is dominated by Brownian motion, the amplitudes are unbounded, and no stationary state occurs.
In particular, these observations confirm again that the conservation of normalisation is a crucial feature of the DNSE, and any violation of such a
conservation law changes the character of the dynamics considerably.

%% The Appendices part is started with the command \appendix;
%% appendix sections are then done as normal sections
%\appendix
%\section{Example Appendix Section}
%\label{app1}

\section{Conclusion} \label{sec:4}

We have summarised aspects of localised solutions of the deterministic
and stochastic DNSE.  Our approach has focused on the impact
of the on-site nonlinearity on the dynamics. We constrained the time
evolution to states which are normalised to $1$. Such a choice is
suitable to study systems of finite size but it misses occasionally aspects
which become vital in the thermodynamic limit of the model. In that sense
our investigations supplement studies of the DNSE which focus on the
statistical mechanics of the model.

If one compares our outcomes with results obtained with those of a statistical mechanics setting, then the investigations of the deterministic DNSE
for $\alpha>0$, \eqref{eqofmotion}, correspond to the study of the standardised model, \eqref{original}, when one approaches temperature zero from below.
Similarly, the dynamically conjugate case of the deterministic DNSE
for $\alpha<0$ corresponds to a study of the standardised DNSE when
temperature zero is approached from above. Both such cases are of course dominated by breather solutions described in the literature and in section \ref{sec:2}. As an aside, one may effectively change the sign of the
temperature by changing the sign of the nonlinear term
in \eqref{eqofmotion}.
The stochastic model with multiplicative noise,
\eqref{eq3.1}, develops
a stationary state which corresponds to the infinite temperature case of
the DNSE. This stochastic model approaches the deterministic case in the limit
of small to vanishing noise level in a very peculiar way. While the deterministic
DNSE has constant energy, the multiplicative noise model shows a constant variance
for the energy which does not depend on the strength of the noise. The
stochastic model approaches the deterministic limit by keeping the variance 
of the energy fixed, but displaying a divergence of the
correlation time in the energy autocorrelation function.
Hence this peculiar limit is singular. The DNSE with additive noise is a fairly
meaningless stochastic model which violates both conservation laws, the conservation of energy and the conservation of normalisation, and thus does not
develop any stationary state. This observation emphasises again the importance
of nontrivial conserved quantities for the dynamical properties of the DNSE.
%Having said that, we do not insinuate that all additive noise models are per se meaningless. 
However, the runaway behaviour of energy and normalization in the model with additive noise can be avoided if one adds suitable dissipation terms. 
%then 
%and 
Then, one can set up a stochastic model 
that naturally leads to a stationary state
representing a grandcanonical equilibrium at finite temperature
as convincingly
demonstrated in \cite{Iubini_2013b}.
These considerations have been extended
recently to cover negative-temperature regimes as well
\cite{iubini_2025}.

In our investigation of the DNSE we have used an approach which allows us
to switch easily between the positive and the negative temperature regime by
changing the sign of the nonlinear term. While the results reported here can be found to some extent in the existing literature in various disguises,
we have aimed at a comprehensible account which is accessible for a large audience.
%\bd{Maybe add s.th. positive about our paper: is it more pedagogical? Is the language or presentation  more understandable for physicists?  Does it bring together results that so far can only be found in n different papers? Do we place things within a wider framework?} 
We have covered in our exposition two
limiting cases, the DNSE for vanishing and for infinite
temperature. The more interesting finite temperature cases could be realised by coupling the DNSE to a heat bath. Studying the resulting stochastic model which now includes noise and damping in line with the required dissipation fluctuation relations could give better insight into stationary 
and dynamical properties of the DNSE, including the negative temperature regime.
%In that sense our studies so far are preliminary and further details will be published elsewhere.
%\bd{again, I find the last sentence (which I commented out) too negative and thought that it invites reviewers to tell us to wait until we have those more advanced results. What about the following: } 
Our present study prepares the ground for such an investigation. 

\section*{Acknowledgements}
We acknowledge useful discussions with Marco Hofmann
 during the early stages of this work, who did preliminary investigations of the model during his Master thesis.

\section*{Funding and Competing Interests}
 This work was supported by the German Research Foundation (DFG) under contract number Dr300/16 and via CRC 1270 ``Electrically Active Implants'', Grant/Award Number SFB 1270/2-299150580.
 
\appendix
\section{Symplectic integrators}\label{sec:a}

Numerical integration of Hamiltonian dynamics over large time intervals require integration schemes which preserve the energy, or which preserve a quantity which
differs little from the energy, so called symplectic integration schemes. Here, we require integration schemes which preserve in addition the normalisation $N$. Following the basic ideas of \cite{Yosh_PLA90}, we design second-order schemes for the autonomous and the non-autonomous case.

\subsection{Autonomous system} \label{sec:a.1}

To design a symplectic integration scheme for eq.~\eqref{eqofmotion}) we write the
Hamiltonian eq.~\eqref{Hamiltonian}) as a sum of two parts, $H=H_A+H_B$ with
\begin{eqnarray}\label{eqa.1}
H_A &=& \sum_{n=0}^{L-1}\left(2 |c_n|^2-c_n \bar{c}_{n-1}-c_n \bar{c}_{n+1}\right)
\nonumber \\   
H_B &=& -\frac{\alpha}{2} \sum_{n=0}^{L-1} |c_n|^4 
 \, .
\end{eqnarray}
Each of these two parts gives rise to a motion that can be computed by analytic means.

If we use $i\mathbf{L}_B |c_n|^2=\{H_B, |c_n|^2\}=0$ and 
$i\mathbf{L}_B c_n=-i \alpha |c_n|^2 c_n$ we obtain for the evolution operator of
$H_B$
\begin{equation}\label{eqa.2}
\exp\left( i \mathbf{L}_B \tau \right) c_n = \exp\left(-i \alpha |c_n|^2 \tau\right)
c_n \, .
\end{equation}

Likewise, we can handle the dynamics of the Hamiltonian $H_A$ which constitutes 
the nearest neighbour coupled chain of harmonic oscillators. If we introduce Fourier modes by
\begin{equation}\label{eqa.3}
\hat{c}_q = \sum_{n=0}^{L-1} \exp(i q n) c_n, \quad q=2\pi \nu/L, \quad
\nu=0,1,\ldots,L-1
\end{equation}
we have
\begin{align}\label{eqa.4}
i\mathbf{L}_A \hat{c}_q &= i\sum_{n=0}^{L-1} \exp(i qn)
\left(2 c_n -c_{n-1}-c_{n+1}\right)\nonumber\\
&=i 2(1-\cos(q)) \hat{c}_q
\end{align}
so that
\begin{equation}\label{eqa.5}
\exp\left(i \mathbf{L}_A \tau \right)\hat{c}_q=
\exp\left(i 2(1-\cos(q))\tau\right) \hat{c}_q \, .
\end{equation}
Thus using the Fourier transform eq.~\eqref{eqa.3}) and its inverse the evolution operator
of $H_A$ can be written in closed form
\begin{align}\label{eqa.6}
\exp\left(i \mathbf{L}_A \tau \right) c_n &= \frac{1}{L} \sum_q
\exp(-i q n) \exp\left(i 2(1-\cos(q))\tau \right) \nonumber\\
&\sum_{m=0}^{L-1} \exp\left(i m q\right) 
c_m \, .
\end{align}
To evaluate the sums in eq.~\eqref{eqa.6}) efficiently, one first computes the Fourier
transform of the state $\{c_n\}$ (see eq.~\eqref{eqa.3})), evaluates then the time
evolution of each Fourier mode by eq.~\eqref{eqa.5}), and finally performs an inverse
Fourier transform.

If $\tau$ denotes the stepsize of a numerical integration scheme, we can approximate the
exact evolution operator of eq.~\eqref{eqofmotion}), 
$\exp(i(\mathbf{L}_A+\mathbf{L}_B)\tau)$, up to second order by
\begin{align}\label{eqa.7}
\exp\left(i(\mathbf{L}_A+\mathbf{L}_B)\tau\right) =
&\exp\left(i \mathbf{L}_B \tau/2\right)
\exp\left(i \mathbf{L}_A \tau\right)\nonumber\\
&\exp\left(i \mathbf{L}_B \tau/2\right) + \mathcal{O}(\tau^3)
\end{align}
as can be straightforwardly verified by using the series expansion of the exponentials.
The right-hand side provides a symplectic integration scheme with a one-step error
of order $\mathcal{O}(\tau^3)$ since each of the three factors can be evaluated
up to machine precision. (i) One first applies eq.~\eqref{eqa.2}) to a state
$\{c_n\}$ with time step $\tau/2$ to obtain an intermediate result. (ii) One then applies 
eq.~\eqref{eqa.6}) to this intermediate result, using a Fourier transform, 
eq.~\eqref{eqa.5}), and an inverse Fourier transform. (iii) Finally one applies eq.~\eqref{eqa.2}) 
again with time $\tau/2$ to obtain the final state $\{c_n\}$ of this symplectic 
integration step.
By design, the scheme preserves an energy which differs from 
eq.~\eqref{Hamiltonian}) by an amount of order $\mathcal{O}(\tau^2)$
(see figure \ref{fig3}(b)). In addition, the normalisation is preserved to
machine precision (see figure \ref{fig3}(c)), since $i\mathbf{L}_A N=0$
and $i\mathbf{L}_BN=0$.

\subsection{Stochastic systems} \label{sec:a.2}

To design a symplectic integration step for stochastic systems, let us first
consider the case of a non-autonomous Hamiltonian system, where the Hamiltonian can be split into two parts, $H(t)=H_A+H_B(t)$, and where we assume for simplicity that
$\{H_B(t),H_B(t')\}=0$. The evolution operator of the non-autonomous system obeys
\begin{equation}\label{eqa.8}
\frac{\partial \mathbf{U}(t_0,t)}{\partial t} = \mathbf{U}(t_0,t) i \mathbf{L}(t),
\quad \mathbf{U}(t_0,t_0)=\mathbf{1} \, .
\end{equation}
The Neumann series up to terms of second order can be easily obtained by iteration
\begin{eqnarray}\label{eqa.9}
\mathbf{U}(t_0,t_0+\tau)&=&\mathbf{1}+\int_{t_0}^{t_0+\tau}
\mathbf{U}(t_0,t) i \mathbf{L}(t) dt \nonumber \\
&=& \mathbf{1}+i\mathbf{L}_A \tau +i \int_{t_0}^{t_0+\tau} \mathbf{L}_B(t) dt \nonumber \\
& & +\int_{t_0}^{t_0+\tau} \int_{t_0}^t \left(i \mathbf{L}_A+i \mathbf{L}_B(s)\right)\nonumber\\
& &\left(i \mathbf{L}_A+i \mathbf{L}_B(t)\right) ds dt +\mathcal{O}(\tau^3)
\nonumber \\
&=& \mathbf{1}+ i\mathbf{L}_A \tau + \left(i\mathbf{L}_B^{<}+ i\mathbf{L}_B^{>} \right)
\tau \nonumber \\
& & + \left(i\mathbf{L}_A\right)^2 \frac{\tau^2}{2}
+ i\mathbf{L}_A i \mathbf{L}_B^{>} \tau^2+
i \mathbf{L}_B^< i \mathbf{L}_A \tau^2 \nonumber \\
& & +\left(i\mathbf{L}_B^<+ i\mathbf{L}_B^>\right)^2
\frac{\tau^2}{2}+ \mathcal{O}(\tau^3)
\end{eqnarray}
where we have introduced the abbreviations
\begin{eqnarray}\label{eqa.10}
i \mathbf{L}_B^< &=& \tau^{-2} \int_{t_0}^{t_0+\tau} 
\int_{t_0}^t i \mathbf{L}_B(s) ds dt\nonumber\\
&=& \tau^{-2} \int_{t_0}^{t_0+\tau} (t_0+\tau-s) i \mathbf{L}_B(s) ds \nonumber \\
i \mathbf{L}_B^> &=& \tau^{-2} \int_{t_0}^{t_0+\tau} 
\int_{t_0}^t i \mathbf{L}_B(t) ds dt\nonumber\\
&=& \tau^{-2} \int_{t_0}^{t_0+\tau} (t-t_0) i \mathbf{L}_B(t) dt \, .
\end{eqnarray}
eq.~\eqref{eqa.9}) can be written as a product
of suitable exponentials, namely (cf. eq.~\eqref{eqa.7}))
\begin{align}\label{eqa.11}
\mathbf{U}(t_0,t_0+\tau) =
&\exp\left(i\mathbf{L}_B^< \tau\right)
\exp\left( i \mathbf{L}_A \tau \right)
\exp\left(i\mathbf{L}_B^> \tau\right)\nonumber\\
&+ \mathcal{O}(\tau^3) \, .
\end{align}
We will base our symplectic integration step on such an identity.
This means that we are implementing noise in the sense of Stratonovich, who treats the stochastic term as the limit of a continuous function. 

{\em i)}:
For the SDNSE with multiplicative noise, eq.~\eqref{eq3.1}), we use
\begin{eqnarray}\label{eqa.12}
H_A &=& \sum_{n=0}^{L-1}\left(2 |c_n|^2-c_n \bar{c}_{n-1}-c_n \bar{c}_{n+1}\right)
\nonumber \\
H_B(t) &=&  \sum_{n=0}^{L-1} 
\left(
-\frac{\alpha}{2} |c_n|^4 +\sigma |c_n|^2 \xi_n(t)\right) 
\end{eqnarray}
to write the Hamiltonian eq.~\eqref{eq3.3}) as a sum of two parts
$H(t)=H_A+H_B(t)$. Obviously the time-dependent part obeys
$\{H_B(t),H_B(t')\}=0$. Eq.~\eqref{eqa.10}) yields for the
effective Hamiltonians $H_B^<$ and $H_B^>$
\begin{equation}\label{eqa.13}
H_B^\lessgtr= \frac{1}{2} \sum_{n=0}^{L-1} 
\left( -\frac{\alpha}{2}\right) |c_n|^4
+ \sum_{n=0 }^{L-1} \sigma |c_n|^2 \frac{D_n^\lessgtr}{\sqrt{\tau}}
\end{equation}
where we have introduced the Gaussian random variables $D_n^\lessgtr$ by
\begin{eqnarray}\label{eqa.14}
D_n^< &=& \tau^{-3/2} \int_{t_0}^{t_0+\tau} (t_0+\tau-t) \xi_n(t) dt \nonumber \\
D_n^> &=& \tau^{-3/2} \int_{t_0}^{t_0+\tau} (t-t_0) \xi_n(t) dt  \, .
\end{eqnarray}
For the correlations of these random variables, we obtain with the help of eq.~\eqref{eq3.2})
\begin{eqnarray}\label{eqa.15}
\langle D_n^< D_m^<\rangle &=& \delta_{n,m} \tau^{-3} \int_{t_0}^{t_0+\tau}
(t_0+\tau-t)^2 d t = \frac{1}{3} \delta_{n,m} \nonumber \\
\langle D_n^> D_m^>\rangle &=& \delta_{n,m} \tau^{-3} \int_{t_0}^{t_0+\tau}
(t-t_0)^2 d t = \frac{1}{3} \delta_{n,m} \nonumber \\
\langle D_n^< D_m^>\rangle &=& \delta_{n,m}  \tau^{-3} \int_{t_0}^{t_0+\tau}
(t_0+\tau-t)(t-t_0) d t \nonumber\\&=& \frac{1}{6} \delta_{n,m} \, .
\end{eqnarray}
eq.~\eqref{eqa.13}) results in $i\mathbf{L}_B^\lessgtr |c_n|^2=0$ and
$i \mathbf{L}_B^\lessgtr=-i \alpha |c_n|^2 c_n/2 + i \sigma D_n^\lessgtr/\sqrt{ \tau}$
so that (cf. eq.~\eqref{eqa.2}))
\begin{equation}\label{eqa.16}
\exp\left(i \mathbf{L}_B^\lessgtr \tau \right) c_n
=\exp\left(-i \alpha |c_n|^2 \tau/2 + i \sigma D_n^\lessgtr \sqrt{\tau}\right)
c_n \, .
\end{equation}
For the action of $\exp(i\mathbf{L}_A \tau)$ we can simply refer to the previous
section \ref{sec:a.1}, eqs.(\ref{eqa.3})-(\ref{eqa.6}).

We now apply eq.~\eqref{eqa.11}) to generate a symplectic integration step
for the SDNSE, eq.~\eqref{eq3.1}), keeping in mind that the stochastic
one-step error is of order
$\mathcal{O}(\tau^2)$ due to the lack of smoothness of random functions.
Given a state $\{c_n\}$ one generates correlated random numbers
$D_n^\lessgtr$ from suitable linear combinations of
uncorrelated normal random variables. (i) First\footnote{Recall that 
$\exp(i\mathbf{L}_{a/b} \tau) c=F_{a/b}(c)$ implies 
$\exp(i \mathbf{L}_a\tau) \exp(\mathbf{L}_b \tau) c =
\exp(i \mathbf{L}_a\tau) F_b(c)= F_b(F_a(c))$.} one applies
eq.~\eqref{eqa.16}) with random variables $D_n^<$ to generate an
intermediate state. 
%\bd{I found the footnote confusing for two reasons: first, I did not understand %that 1/2 does not mean 0.5 but 1 or 2, and second I think that F2(F1(c)) must be %the other way round, i.e. F1(F2(c))} 
(ii) Then one performs a Fourier transform, 
applies eq.~\eqref{eqa.5}), and performs an inverse Fourier transform. 
(iii) Finally, we use again eq.~\eqref{eqa.16}) with random variables
$D_n^>$ to obtain the final state of the numerical integration scheme.
Since $i\mathbf{L}_A N=0$ and $i\mathbf{L}_B^\lessgtr N=0$ the numerical
scheme preserves the normalisation.

{\em ii)}: For the SDNSE with additive noise, eq.~\eqref{eq3.4}), we use
\begin{eqnarray}\label{eqa.17}
H_A &=& \sum_{n=0}^{L-1}\left(2 |c_n|^2-c_n \bar{c}_{n-1}-c_n \bar{c}_{n+1}
-\frac{\alpha}{2} |c_n|^4\right)
\nonumber \\
H_B(t) &=& \sigma \sum_{n=0}^{L-1} 
\left(c_n+\bar{c}_n \right) \xi_n(t) 
\end{eqnarray}
to write the Hamiltonian eq.~\eqref{eq3.5}) as a sum of two parts
$H(t)=H_A+H_B(t)$. Obviously the time-dependent part obeys
$\{H_B(t),H_B(t')\}=0$. Eq.~\eqref{eqa.10}) yields for the
effective Hamiltonians $H_B^<$ and $H_B^>$
\begin{equation}\label{eqa.18}
H_B^\lessgtr= \sigma \sum_{n=0}^{L-1} 
\left(c_n + \bar{c}_n\right) \frac{D_n^\lessgtr}{\sqrt{\tau}}
\end{equation}
where $D_n^\lessgtr$ again denote the random variables defined in eq.~\eqref{eqa.14})
and (\ref{eqa.15}). Since $i\mathbf{L}_B^\lessgtr c_n=i \sigma D_n^\lessgtr/\sqrt{\tau}$
we have (cf. eq.~\eqref{eqa.16}))
\begin{equation}\label{eqa.19}
\exp\left( i \mathbf{L}_B^\lessgtr \tau\right) c_n= c_n+ i \sigma D_n^\lessgtr \sqrt{\tau} \, .
\end{equation}
Since $H_A$ in eq.~\eqref{eqa.17}) coincides with the Hamiltonian of the
DNSE, eq.~\eqref{Hamiltonian})
we can evaluate $\exp(i\mathbf{L}_A \tau)$ (up to and including second order
terms) by the scheme described in section \ref{sec:a.1}, cf. eq.~\eqref{eqa.7}).

We now use again eq.~\eqref{eqa.11}) to design a symplectic integration step with
one-step error of order $\mathcal{O}(\tau^2)$. 
Given a state $\{c_n\}$ one generates correlated random numbers
$D_n^\lessgtr$ from suitable linear combinations of
uncorrelated normal random variables. (i) The first one applies
eq.~\eqref{eqa.19}) with random variables $D_n^<$ to generate an
intermediate state. (ii) Then, following the steps of section \ref{sec:a.1}
one applies eq.~\eqref{eqa.2}) with time step $\tau/2$ 
to the intermediate state, performs a Fourier transform,
applies eq.~\eqref{eqa.5}) to the Fourier
components, performs an inverse Fourier transform, and applies eq.~\eqref{eqa.2}) 
again with time $\tau/2$. (iii) Finally we use eq.~\eqref{eqa.19}) with 
random variables $D_n^>$ 
to obtain the result of the symplectic integration step.

% The \nocite command causes all entries in a bibliography to be printed out
% whether or not they are actually referenced in the text. This is appropriate
% for the sample file to show the different styles of references, but authors
% most likely will not want to use it.

%% \nocite{*}

%Appendix text.

%% For citations use: 
%%       \citet{<label>} ==> Lamport [21]
%%       \citep{<label>} ==> [21]
%%
%Example citation, See \citet{lamport94}.

%% If you have bib database file and want bibtex to generate the
%% bibitems, please use
%%
%%  \bibliographystyle{elsarticle-num-names} 
%%  \bibliography{<your bibdatabase>}

%% else use the following coding to input the bibitems directly in the
%% TeX file.

%% Refer following link for more details about bibliography and citations.
%% https://en.wikibooks.org/wiki/LaTeX/Bibliography_Management

%\begin{thebibliography}{00}

%% For authoryear reference style
%% \bibitem[Author(year)]{label}
%% Text of bibliographic item

%\bibitem[Lamport(1994)]{lamport94}
%  Leslie Lamport,
%  \textit{\LaTeX: a document preparation system},
%  Addison Wesley, Massachusetts,
%  2nd edition,
%  1994.

%%\bibliographystyle{elsarticle-harv} 
%%\bibliography{apssamp}

\providecommand{\noopsort}[1]{}\providecommand{\singleletter}[1]{#1}%

%\end{thebibliography}
\end{document}